\documentclass[
 amsmath,amssymb,
 aps,prd,twocolumn,
 footinbib,preprintnumbers
]{revtex4}
\usepackage{lipsum}
\usepackage{graphicx}
\usepackage{dcolumn}
\usepackage{xcolor}
\usepackage[detect-all]{siunitx}
\usepackage{bm}
\usepackage{braket}
\usepackage{tikz}
\usepackage{mathtools}
\usepackage{mathtools}

\makeatletter
\ifx\@ptsize\undefined\def\@ptsize{0}\fi
\makeatother
\usepackage{amsmath, amssymb, graphics, setspace}

\newcommand{\mathsym}[1]{{}}
\newcommand{\unicode}[1]{{}}

\usepackage{graphicx}
\graphicspath{{/}}
\usepackage{dcolumn}
\usepackage{bm}

\usepackage[unicode,hidelinks,linktocpage,colorlinks=true,
            plainpages=false,pdfpagelabels,hyperfootnotes=false]{hyperref}
\hypersetup{urlcolor={blue},citecolor={black},menucolor={black},linkcolor={black}}
\ifx\onlinecite\undefined
\ifx\inlinecite\undefined\def\onlinecite{\cite}\else
\def\onlinecite{\inlinecite}\fi\fi
\ifx\inlinecite\undefined\def\inlinecite{\onlinecite}\fi

\usepackage[capitalize]{cleveref}

{\catcode`\/=12\relax
\gdef\makesplittable#1/#2\endsplittable{%
  \dodotsplittable{#1}\ifx\relax#2\relax\else/\discretionary{}{}{}%
  \makesplittable#2\endsplittable\fi}
\gdef\dosplittable#1{{\makesplittable#1/\endsplittable}}}
{\catcode`\.=12\relax
\gdef\makedotsplittable#1.#2\endsplittable{%
  \dolparensplittable{#1}\ifx\relax#2\relax\else.\discretionary{}{}{}%
  \makedotsplittable#2\endsplittable\fi}
\gdef\dodotsplittable#1{{\makedotsplittable#1.\endsplittable}}}
{\catcode`\(=12\relax
\gdef\makelparensplittable#1(#2\endsplittable{%
  \dorparensplittable{#1}\ifx\relax#2\relax\else\discretionary{}{}{}(%
  \makelparensplittable#2\endsplittable\fi}
\gdef\dolparensplittable#1{{\makelparensplittable#1(\endsplittable}}}
{\catcode`\)=12\relax
\gdef\makerparensplittable#1)#2\endsplittable{%
  #1\ifx\relax#2\relax\else)\discretionary{}{}{}%
  \makerparensplittable#2\endsplittable\fi}
\gdef\dorparensplittable#1{{\makerparensplittable#1)\endsplittable}}}
\newcommand\dourl[1]{{\edef\temp{\noexpand\dosplittable{#1}}\expandafter}\temp}
\newcommand\dodoi[1]{{\edef\temp{\noexpand\dosplittable{#1}}\expandafter}\temp}

\allowdisplaybreaks

\begin{document}

\preprint{LA-UR-20-24538}

\ifx\texorpdfstring\undefined\def\texorpdfstring#1#2{#1}\fi

\title{State preparation and measurement\texorpdfstring{\\}{ }in a quantum simulation of the $O(3)$ sigma model}

\author{Alexander J. Buser}
\affiliation{Caltech, Institute for Quantum Information
and Matter, Pasadena, CA 91125}
\author{Tanmoy Bhattacharya}
\affiliation{Los Alamos National Laboratory, Theoretical Division, Los Alamos, NM 87545}
\author{Lukasz Cincio}
\affiliation{Los Alamos National Laboratory, Theoretical Division, Los Alamos, NM 87545}
\author{Rajan Gupta}
\affiliation{Los Alamos National Laboratory, Theoretical Division, Los Alamos, NM 87545}
\date{\today}

\begin{abstract}
Recently, Singh and Chandrasekharan~\cite{PhysRevD.100.054505} showed that fixed points of the non-linear $O(3)$ sigma model can be reproduced near a quantum phase transition of a spin model with just two qubits per lattice site. In a paper by the NuQS collaboration~\cite{PhysRevLett.123.090501}, the proposal is made to simulate such field theories on a quantum computer using the universal properties of a similar model. In this paper, following that direction, we demonstrate how to prepare the ground state of the model from~\cite{PhysRevD.100.054505} and measure a dynamical quantity of interest, the $O(3)$ Noether charge, on a quantum computer. In particular, we apply Trotter~\cite{Trotter} methods to obtain results for the complexity of adiabatic ground state preparation in both the weak-coupling and quantum-critical regimes and use shadow tomography~\cite{2002.08953} to measure the dynamics of local observables. We then present and analyze a quantum algorithm based on non-unitary randomized simulation methods that may yield an approach suitable for intermediate-term noisy quantum devices.
\end{abstract}

\maketitle

\section{\label{sec:level1}Introduction}
Considering the rapid pace of experimental progress in quantum computing, now is the time to think seriously about how quantum simulations of high energy physics will look in the near future. Quantum field theories generally involve infinitely many degrees of freedom per unit physical volume. A crucial step in simulating field theories, then, is choosing a method of truncating the Hilbert space and representing the states of the truncated model using qubits. According to the standard procedure advocated by Wilson~\cite{Wilson:1974sk}, one regulates the theory via a hard cutoff, restricting it to a discrete finite lattice. Bosonic theories, however, involve infinitely many quantum states per lattice site even after regularization. Realizing the theory as a quantum simulation on a finite computer requires us to truncate the dimension of the Hilbert space, being careful to preserve the interesting physics of the theory.

One approach is to impose a hard cut-off on the occupation number in the Fourier space of the relevant symmetry group, and directly map the truncated Hamiltonian to one acting on a system of qubits~\cite{PhysRevA.73.022328}. This approach is provably efficient for Kogut-Susskind type lattice gauge theories~\cite{Kogut:1974ag}, and trivially converges to the correct behavior as the local dimension is scaled back to infinity. In the near-term, however, one will be limited to small local dimension with only few qubits per lattice site. Such truncated models are not, in general, good approximations to the original field theory.

We therefore advocate working directly with the standard definition of the continuum field theory as describing the long distance physics near the critical points of a discretized theory. This is particularly interesting for asymptotically-free theories that arise near Gaussian fixed points of the renormalization group evolution. To this end, we construct a lattice Hamiltonian possessing the same symmetries as the desired continuum QFT, but acting on a Hilbert space with small local dimension. This is the approach taken in Refs.~\onlinecite{PhysRevD.100.054505,PhysRevLett.123.090501} and extended in \cref{sec:general} for the case of the $O(3)$ non-linear sigma model. 

The present paper is concerned with the second step of this process---having found a lattice model in the right universality class, how does one go about simulating it on a quantum computer? In Ref.~\onlinecite{PhysRevLett.123.090501}, the authors present an efficient circuit for performing time-evolution on a model with two qubits per lattice site using just $12$ CNOT gates per step of Trotterized time evolution. Since the model we consider is distinct from the fuzzy sphere discretization introduced in Ref.~\onlinecite{PhysRevLett.123.090501}, we do not use their circuit directly, but construct one for our model using symmetry arguments. This circuit allows us to derive bounds on the complexity of interesting computational tasks, although our approach likely will not be implemented on a near-term device due to the large number of two-qubit gates required. Instead, we apply quantum circuit compiling to circumvent the large resource requirements. We describe the implementation of two interesting and non-trivial tasks, namely ground state preparation and the measurement of dynamic quantities, in particular the $O(3)$ Noether current, and provide estimates of the complexity of each task. Of particular interest to the quantum computing community is an algorithm developed here combining quantum circuit compiling and randomized simulation methods for adiabatic state preparation, and an application of shadow tomography to measuring local observables in quantum field theory.

This paper is organized as follows. In \cref{sec:alg} we describe the model and discuss the preparation of the ground state. In \cref{sec:Noether} the measurement of the $O(3)$ Noether charge is discussed. In \cref{sec:err} we estimate the resources required for ground state preparation in both the weak-coupling and quantum-critical regimes. In \cref{sec:nisq} we
% use classically simulated quantum-assisted quantum compiling (QAQC)~\cite{Khatri2019quantumassisted} in conjunction with a randomized simulation algorithm~\cite{randomized} to
find short-depth circuits approximating the adiabatic state-preparation algorithm and study them numerically. We end with a discussion of our main results in \cref{sec:conc}.  Technical details are provided in appendices.

\section{\label{sec:alg}\label{sec:CI}Exact algorithm}

In this section, we describe the model studied in Ref.~\onlinecite{PhysRevD.100.054505} using a slightly more convenient convention. We then describe a quantum circuit for preparing the ground state of this system using the Trotter approximation~\cite{Trotter} on a fault-tolerant quantum computer and numerically study the fidelity of the process. We then discuss the resource requirements for this algorithm.

\subsection{The model}
Singh and Chandrasekharan~\cite{PhysRevD.100.054505} recently demonstrated that, using just two qubits per lattice site, a direct truncation scheme reproduces the Wilson-Fisher fixed point of the $O(3)$ sigma model in two spatial dimensions and the Gaussian fixed point in three spatial dimensions \footnote{This method can be easily generalized to the $O(N)$ non-linear sigma model, see Ref.~\onlinecite{Singh:2019jog}.}. The model considered there resides on a regular square or cubic lattice of length $L$ in $d$ spatial dimensions. We describe the system in the basis of the total angular momentum $J$ of two spin-$\frac12$ degrees of freedom representing the two qubits at each site. For the site at position $x$, let $\ket{s,x}$ be the $J=0$ singlet state, and $\ket{m,x}, m=-1,0,+1$ be the $J=1$ triplet states with angular momentum component along the $z$-axis having the value $m$. The Hamiltonian consists of on-site and nearest-neighbor terms.
\begin{equation}\label{eqn:model}
\begin{split}
    H_{\hphantom1} &= H_1 + H_2, \\
    H_1 &= \sum_{x,m} (J + \mu m)\ket{m,x}\bra{m,x}, \\
    H_2 &= J_r(H_p + H_h), \\
    H_p &= -\sum_{\mathclap{\langle x,x'\rangle, m}} (-1)^m\ket{m,x;-m,x'}\bra{s,x;s,x'} + \text{h.c.}, \\
    H_h &= \sum_{\mathclap{\langle x,x'\rangle, m}} \ket{s,x;m,x'}\bra{m,x;s,x'} + \text{h.c.}
\end{split}
\end{equation}
The choice of the sign of the couplings here is slightly different from that in Ref.~\onlinecite{PhysRevD.100.054505}, owing to a different choice of phases for the basis states.

The model contains three independent coupling constants: the on-site coupling $J$ is the extra energy of the triplet states, $\mu$ is the splitting between the triplet states, and $J_r$ is the nearest-neighbor coupling. When $J_r$ and $\mu$ are small and \(J\) is positive, the vacuum is close to the all-singlet state, and one can think of the \(\ket m\) as particle excitations with mass \(J\), \(\mu\) a `chemical potential' corresponding to this particle, \(H_h\) the kinetic term, and \(H_p\) the pair creation/annihilation interaction. 

This model is a qubit-representation of the non-linear $O(3)$ sigma model with a hard cut-off in angular momentum at $l = 1$ (see details in~\cref{sec:reptheory}). In this work, we study only the \(\mu=0\) case, when the theory has a global $O(3)$ symmetry (see details in~\cref{sec:reptheory}), and so conserves total angular momentum $J$ and its $z$-component $M$. The Hamiltonian also has two more symmetries: the number of sites in state $\ket{s}$ modulo 2, and the parity symmetry. We normalize the Hamiltonian by setting $J = 1$, so that the only free parameter is $J_r$, hereafter referred to as the coupling constant. Since \(H_p\) and \(H_h\) separately have all the symmetries of the full model, we need not have the same coefficient for both, so this is actually a choice we are making.

\begin{figure}[t]
    \centering
    \includegraphics[width=150px]{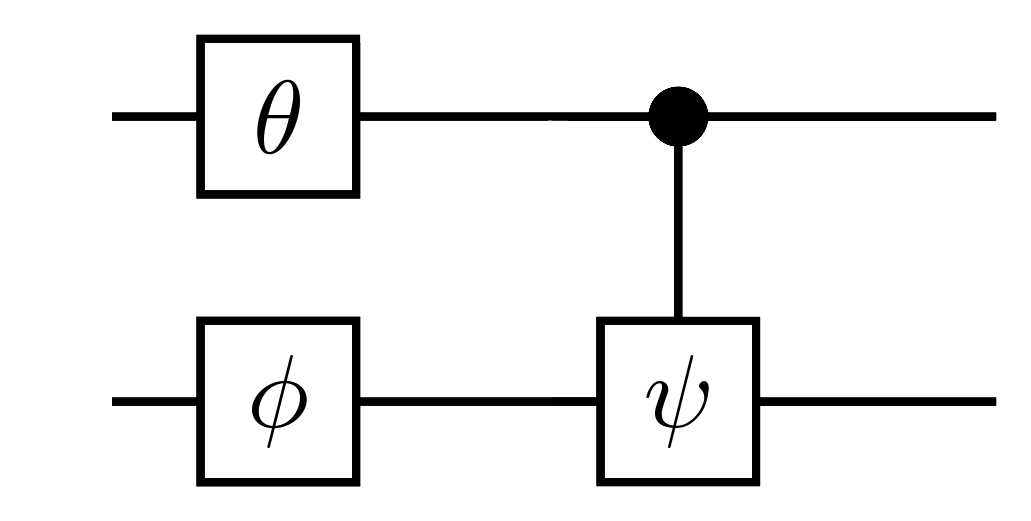}
    \caption{Circuit performing arbitrary diagonal unitary on two qubits. Operators are z-rotations by the corresponding angle. One degree of freedom is set to zero by the global phase, and the other three uniquely determine $\theta$, $\phi$, and $\psi$.}
    \label{fig:1}
\end{figure}

\subsection{Adiabatic state preparation}
\begin{figure*}[t]
    \centering
    \includegraphics[width=230px]{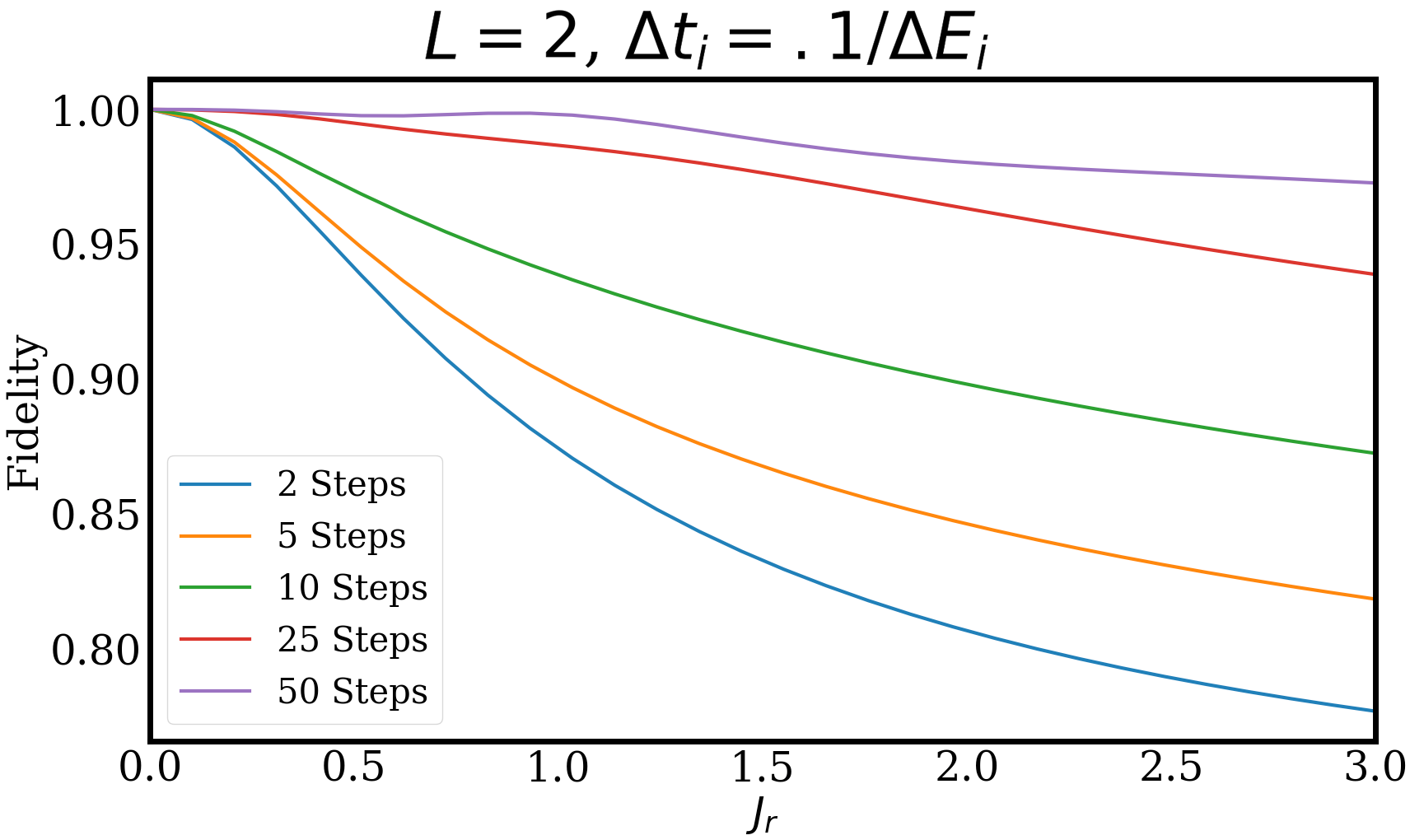}
    \includegraphics[width=230px]{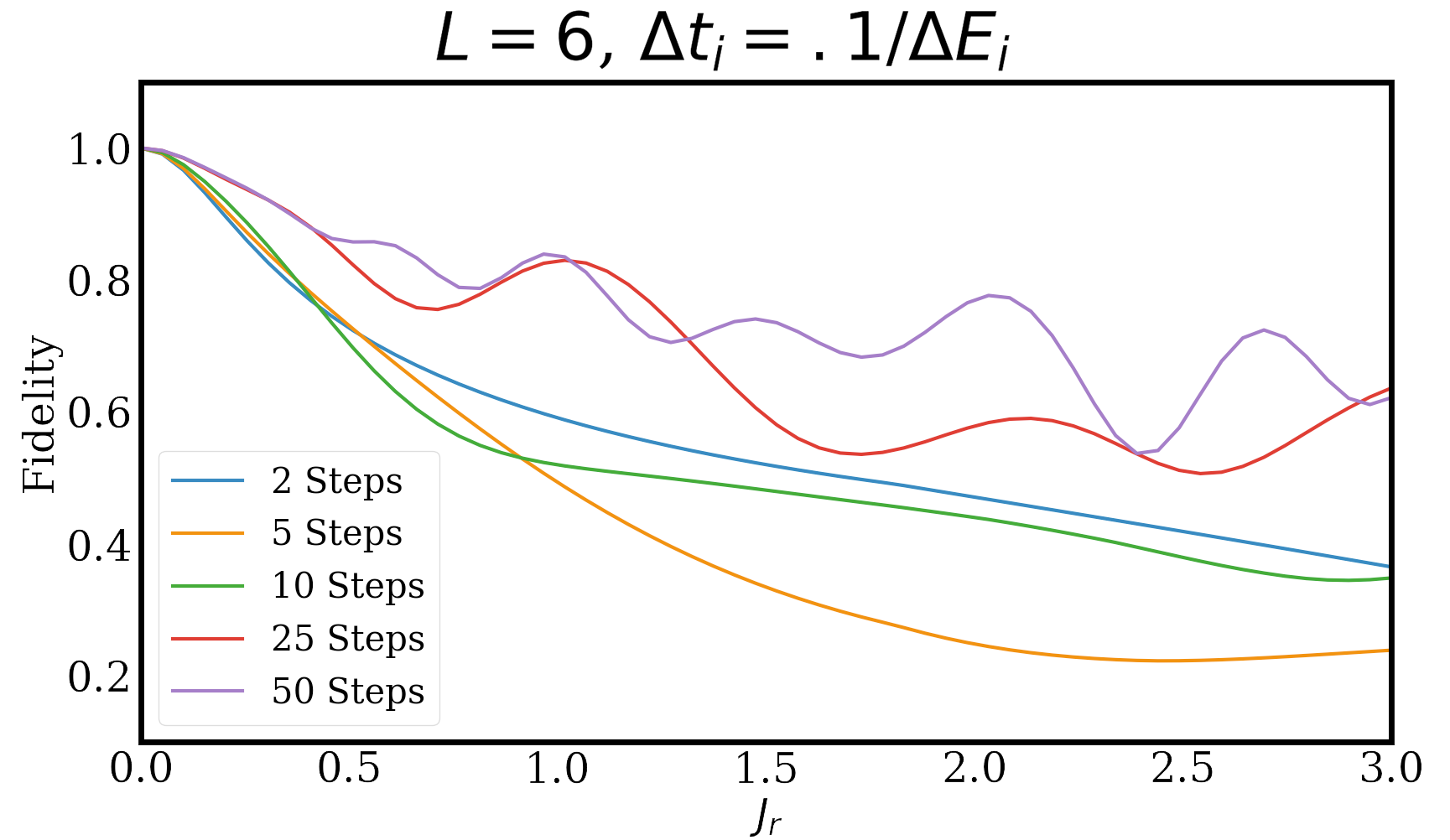}
    \caption{Accuracy of adiabatic state preparation on two- (left) and six- (right) site system as a function of maximum coupling strength $J_r$ for various numbers of Trotter steps. Here, the time step $\Delta t_i$ is chosen to be one-tenth of the inverse of the energy gap, $\Delta E_i$, at coupling $J_{r,i}$. Fidelity between the adiabatically-prepared state $\ket{\psi}$ and the true ground state $\ket{\Omega}$ is defined as $\sqrt{\braket{\psi|\Omega}\braket{\Omega|\psi}}$, where both $\ket{\Omega}$ and $\ket{\psi}$ are properly normalized. Fidelity can be further improved by simultaneously increasing the number of Trotter steps and reducing the time step to minimize non-adiabaticity and Trotterization error.}
    \label{fig:adiabatic2}
    \label{fig:adiabatic6}
    \label{fig:adiabatic}
\end{figure*}
The first task of quantum simulation is to prepare initial states for a given Hamiltonian. The Hilbert space of a quantum field theory consists of the closure of the polynomials of particle creation and annihilation operators acting on the vacuum, i.e., the ground state---so preparing ground states is of fundamental importance. In the presence of an energy gap at all values of the couplings considered, we may prepare the ground state via the adiabatic algorithm~\cite{adiabatic}. Although many algorithms exist for simulating time-evolution with better asymptotic scaling~\cite{7354428,Childs2019fasterquantum,8555119}, we choose to work with standard Trotter methods. This is because the symmetries of the present model make this kind of approach particularly simple to analyze. Furthermore, Trotter formulae are well-suited to the randomized algorithm presented in \cref{sec:nisq}.

The adiabatic algorithm works by choosing a Hamiltonian whose ground state is easy to prepare, and then slowly tuning a set of coupling constants to reach a target Hamiltonian without exciting the system away from the ground state appreciably throughout the whole computation. This requires the evolution to be slow on the time scale of the mass gap of the system. 

Specializing to the model in \cref{eqn:model}, when $J_r = 0$ the ground state is the trivial all-singlet state. We define our computational basis in such a way that, for a pair of qubits at a single site, $\ket{00} = \ket{s}, \ket{01} = \ket{-1}, \ket{10} = \ket{0}, \ket{11} = \ket{1}$. Note that, for instance, what we will refer to as a singlet state does not correspond to the singlet state of two physical qubits, but as long as this choice is made consistently throughout our algorithm, we never need to implement this unitary operation explicitly. Starting from the zero-coupling ground state, which, being a product state, is easy to prepare, we reach the ground state at finite coupling, $\ket{\Omega(J_r)}$, by a series of discrete time-steps;
\begin{equation}\label{eqn:adiabatic}
    \ket{\Omega(J_r)} \approx \prod_{i=1}^N e^{-i \Delta t_i(H_1 + J_{r,i} (H_p + H_h))} \ket{\Omega(0)}.
\end{equation}
Here we have allowed the length of the time-step $\Delta t_i$ and the coupling $J_{r,i}$ to change with each iteration, $i$, and have let $N$ be the number of time-steps. We approximate each time-step by separating out the single-site piece
\begin{align} \label{eqn:approx1}
  e^{-i \Delta t_i(H_1 + J_{r,i} (H_p + H_h))}  &\approx&\nonumber\\
  \omit\span\omit\span\qquad
	e^{-i \Delta t_i H_1} e^{-i\Delta t_i J_{r,i}(H_p + H_h)}
	+ \mathcal{O}(L^d \Delta t ^2).
\end{align}
Since $H_1$ is diagonal in our computational basis this term is easy to implement. As shown in \cref{fig:1}, for a single site $x$,
$\exp \{-i\Delta t_i(J+\mu m)\allowbreak \ket{m,x}\bra{m,x}\}$ may be implemented using just two single-qubit rotations and one controlled rotation even for nonzero $\mu$. Because the adiabatic evolution maintains the translational invariance of the ground state, any possible global phase may be removed consistently from each qubit so that the on-site term is implemented exactly with $3L^d$ gates. 

Next, we decompose the nearest-neighbor term into a product of $2d$ non-commuting operators~\cite{2019arXiv190100564C}. In one dimension, this involves splitting the links of the lattice into two disjoint sets of even and odd links;
\begin{align}\label{eqn:approx2}
  e^{-i\Delta t_i J_{r,i}(H_p + H_h)} &\approx&\nonumber\\
  \omit\span\omit\span\qquad
  e^{-i\Delta t_i J_{r,i}H_{even}}e^{-i\Delta t_i J_{r,i} H_{odd}} + \mathcal{O}(L^d \Delta t^2).
\end{align}
Here $H_{even}$ and $H_{odd}$ denote $H_2$ from \cref{eqn:model} restricted to the sites connected by even and odd links, respectively. This is easily generalized to a cubic lattice in arbitrary dimension. Assuming $L$ is even, each term within $H_{even}$ commutes with all the rest (likewise for $H_{odd}$), since the links are disconnected from each other and $H_h$ and $H_p$ operate on different symmetry sectors (see \cref{sec:general}). Now it suffices to simulate the hopping and pair-creation terms on a single pair of adjacent sites. To do this exactly, we can follow an approach similar to Ref.~\onlinecite{PhysRevA.73.022328} and introduce a unitary operator $U_{CG}$ that implements the Clebsch-Gordan transform on the computational basis. Specifically, this takes a state $\ket{j_1,m_1;j_2,m_2}$ in the local angular momentum basis and gives a state $\ket{J,M,p}$ in the total angular momentum basis \footnote{In order to make this unitary we also need to keep track of the channel $p$ that produced the $\ket{J,M,p}$ state; for example, $J=1$ can arise out of the three channels $\ket{m,x;m,x'}$, $\ket{m,x;s,x'}$ and $\ket{s,x;m,x'}$.}. $H_p$ acts nontrivially only on the $2$-dimensional $J=0$ sector. Similarly, the $H_h$ term acts only on the 9-dimensional $J=1$ sector, decomposing into three \(2\times2\) blocks and one \(3\times3\) diagonal block (See \cref{sec:general}). Thus, evaluating the nearest-neighbor time-step reduces to implementing four two-qubit unitaries. 

For a one-dimensional lattice of small size, the above procedure can be simulated exactly on a classical computer. The accuracy of the output is a complicated function of the adiabatic scheduling, that is, the number of time-steps and the values of $J_r$ and $\Delta t$ at each step. In \cref{fig:adiabatic} we plot the fidelity as a function of the target coupling $J_r$ for five different numbers of time steps to simulate the adiabatic ground state preparation.

For each value of the maximum coupling, we prepare the ground state of our model at $J_r = 0$, and apply the unitary operation in \cref{eqn:adiabatic} under the approximations of \cref{eqn:approx1,eqn:approx2}. The couplings are linearly interpolated between $0$ and the maximum coupling. We require $\Delta t \Delta E \ll 1$ for adiabaticity, where $\Delta E$ is the energy gap of the model, so we set $\Delta t = 0.1/\Delta E$. As expected, the procedure works best for small values of the target coupling, and improves markedly as the number of time steps is increased. For a fixed $J_r$ and number of time steps, however, the fidelity decreases significantly as the lattice size is increased from $2$ to $6$ as in \cref{fig:adiabatic}. This reflects the introduction of Trotter error which increases linearly in the lattice volume, owing to the even-odd decomposition described in \cref{eqn:approx2}.

\subsection{Resource Requirements}

\begin{figure}[t]
    \centering
    \includegraphics[width = \linewidth]{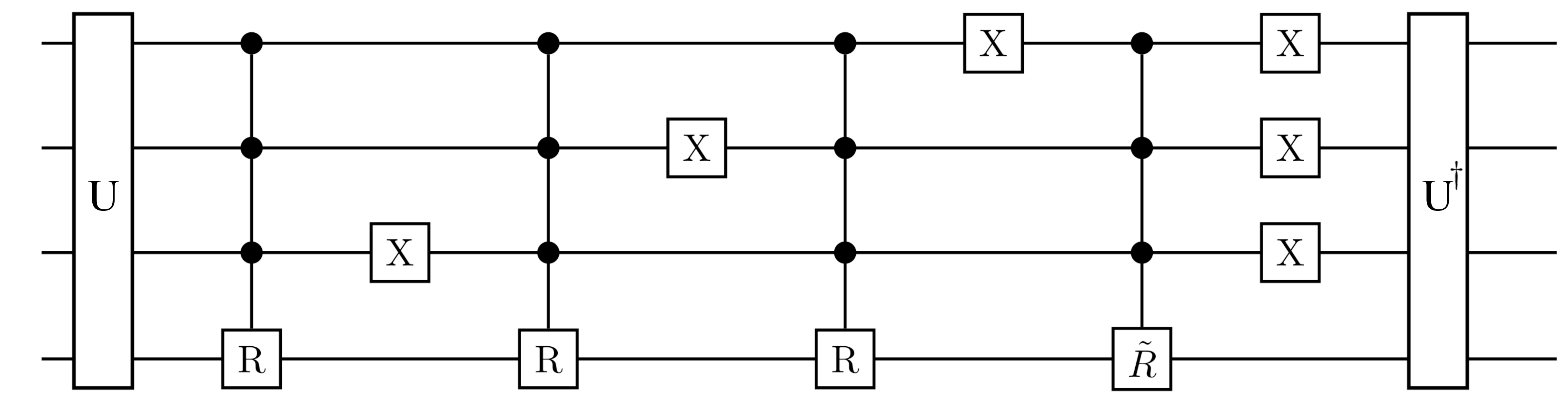}
    \caption{Circuit implementing time evolution of the single-link nearest neighbor term. \(R\) performs the two-level unitary corresponding to the hopping term, while $\tilde{R}$ handles the pair creation. The $U$ gate represents the Clebsch-Gordan transformation on four qubits, schematically shown in \protect\cref{fig:my_label2}, ordering the resulting states such that the controlled two-level gates act on the right states. Using the Sleater-Weinfurter construction~\cite{PhysRevLett.74.4087}, the three-qubit control gates can be rewritten in terms of single-qubit controls.}
    \label{fig:my_label1}
\end{figure}

The number of quantum gates demanded by the algorithm described in the previous subsection makes it impractical to implement on noisy intermediate-scale quantum computers. Even neglecting the cost of the Clebsch-Gordan transformation, the circuit in \cref{fig:my_label1} requires six single qubit gates and four controlled unitaries with three qubit control. In terms of arbitrary single-qubit and single-controlled-unitaries, this requires a total of 58 gates per time step~\cite{PhysRevLett.74.4087, Nielsen:2011:QCQ:1972505} (see \cref{fig:my_label1}).

\begin{figure}[t]
    \centering
    \includegraphics[width=0.9\columnwidth]{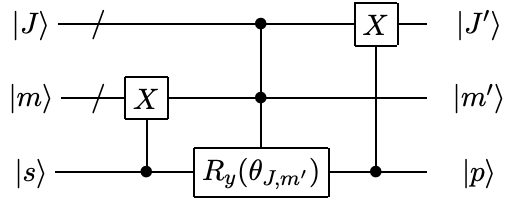}
    \caption{Schematic description of the Clebsch-Gordan transformation by Bacon, Chuang, and Harrow~\cite{PhysRevLett.97.170502}. Here $s$ represents a single qubit which is added to a system with angular momentum $J$ and $m$. On the right, $J'$ and $m'$ label the resulting angular momenta, and $p$ keeps track of the pathway through which those values are obtained. The slash denotes a wire containing a register of qubits, the control-X gate adds angular momentum appropriately, and the controlled rotations (one for each $J$, $m'$) produce the correct amplitudes.}
    \label{fig:my_label2}
\end{figure}

In addition, one must implement the Clebsch-Gordan transformation and its inverse. A schematic description of a circuit implementing this transformation was provided by Bacon, Chuang, and Harrow~\cite{PhysRevLett.97.170502} (see \cref{fig:my_label2}), however it requires ancilla qubits and many-qubit control gates. Although the number of gates required to implement a single time-step this way scales linearly with system size, implementing it exactly seems out of reach on current experimental platforms. In \cref{sec:nisq} we propose a resolution of this problem by finding short-depth circuits that approximate the desired unitary dynamics via numerical optimization. 

%% To calculate the total resource requirements, we also need to calculate the number of time-steps \(T\) required to reach the ground-state to a given accuracy \(\varepsilon\).  This is primarily controlled by the energy gap of the system. The energy gap at $J_r = 0$ is simply $J\equiv1$, but computing it exactly for an arbitrary value of $J_r$ is computationally intractable since it requires time exponential in the number of qubits, so, here, we analyze it in the weak-coupling regime.  Using Rayleigh-Schr\"{o}dinger perturbation theory to compute the energy gap in powers of $J_r$, we find (\cref{sec:perturbative})
%% \begin{equation}
%%     \Delta = 1-2dJ_r + 3dJ_r^2 + \mathcal{O}(J_r^3).
%% \end{equation}
%% Performing adiabatic state preparation using a na\"\i ve first-order Lie product formula, this implies that bounding the total error of adiabatic state preparation to $\varepsilon$ uses a simulation time 
%% \begin{equation}
%%     T = \mathcal{O}\left(\left(\frac{J_{r}^2\mathcal{V}^3}{\varepsilon}\right) + \sqrt{\varepsilon d L^d}\right)
%% \end{equation} 
%% or longer, to leading order in $J_{r}$ (see \cref{sec:err}). 

\section{\label{sec:Noether}Noether Current}
One promising application of quantum computing to high energy physics is the potential to measure dynamic quantities in real-time. This requires both the excitation of interesting initial states out of the vacuum (i.e., ground state) and measurement of the quantities of interest.  In this study, we focus on the latter problem alone.

As a prototypical example of this, we consider the problem of measuring an interesting local observable which is simple to write down, namely the $O(3)$ Noether current. The $z$-component of the Noether charge $Q_{z,x}$ simply counts the angular momentum in the z-direction at the site $x$. Because of the $O(3)$ symmetry of our model, $Q_z = \sum_x Q_{z,x}$ is a conserved quantity. However, there may be interesting physics contained in the local fluctuations of $Q_{z,x}$. To find the current, $J_{z,x}$, of this charge we use the following relation
\begin{equation}
    [Q_{z,x}, H] = -i(\sum_{x'} J_{z,\langle x,x'\rangle})\,
\end{equation}
where $\langle x,x'\rangle$ represent a pair of neighboring sites.
Evaluating this commutator explicitly, we obtain the following two-site operator for $J_{z,\langle x,x'\rangle}$:
\begin{equation}
\begin{split}
    J_{z,\langle x,x'\rangle} = iJ_r \sum_m m (&\ket{s\,m}\bra{m\,s} - \ket{m\,s}\bra{s\,m}  \\
    &+ \ket{s\,s}\bra{m\,{-m}} - \ket{-m\,m}\bra{ss}),
\end{split}
\end{equation}
where for convenience we have suppressed the position labels and used a compressed notation: $\ket{p\,q} \equiv \ket{p,x}\ket{q,x'}$. On a large scale fault-tolerant device, one could efficiently compute the expectation value of this observable using the standard technique of phase estimation~\cite{Kitaev:1995qy}. This would allow one to determine the Noether current at a single site to precision $\varepsilon$ using $\mathcal{O}(\text{log}(1/\varepsilon))$ ancilla qubits and $\mathcal{O}(1/\varepsilon)$ controlled-U gates, where $U = \exp(i J_{z,\langle x,x'\rangle})$, so that the Noether current everywhere can be computed in time $\mathcal{O}(\mathcal{V}/\varepsilon)$, where the number of links is $\mathcal{V} = dL^d$.

For a noisy device with a limited number of qubits, however, the need for a large number of ancilla qubits and a circuit depth linear in system size makes applying full-scale phase estimation somewhat impractical. A recent algorithm by Huang et al. applies a technique known as shadow tomography to predict expectation values of $M$ local observables using measurements on $\text{log}(M)$ copies of a quantum state~\cite{2002.08953,aaronson}. By measuring a quantum state in a sequence of random bases, one can construct an efficient classical representation of the state, which suffices for predicting expectation values of a set of local observables. For a local observable like the Noether current, obtaining a precision $\delta$ uses only $\mathcal{O}(\text{log}\mathcal{V}/\delta^2)$ independent copies of the system.  This protocol, free from the need for ancilla qubits and long coherence times, appears to be more suitable for near-term experiments than phase estimation. 

\begin{figure*}[t]
    \centering
    \includegraphics[height=130px]{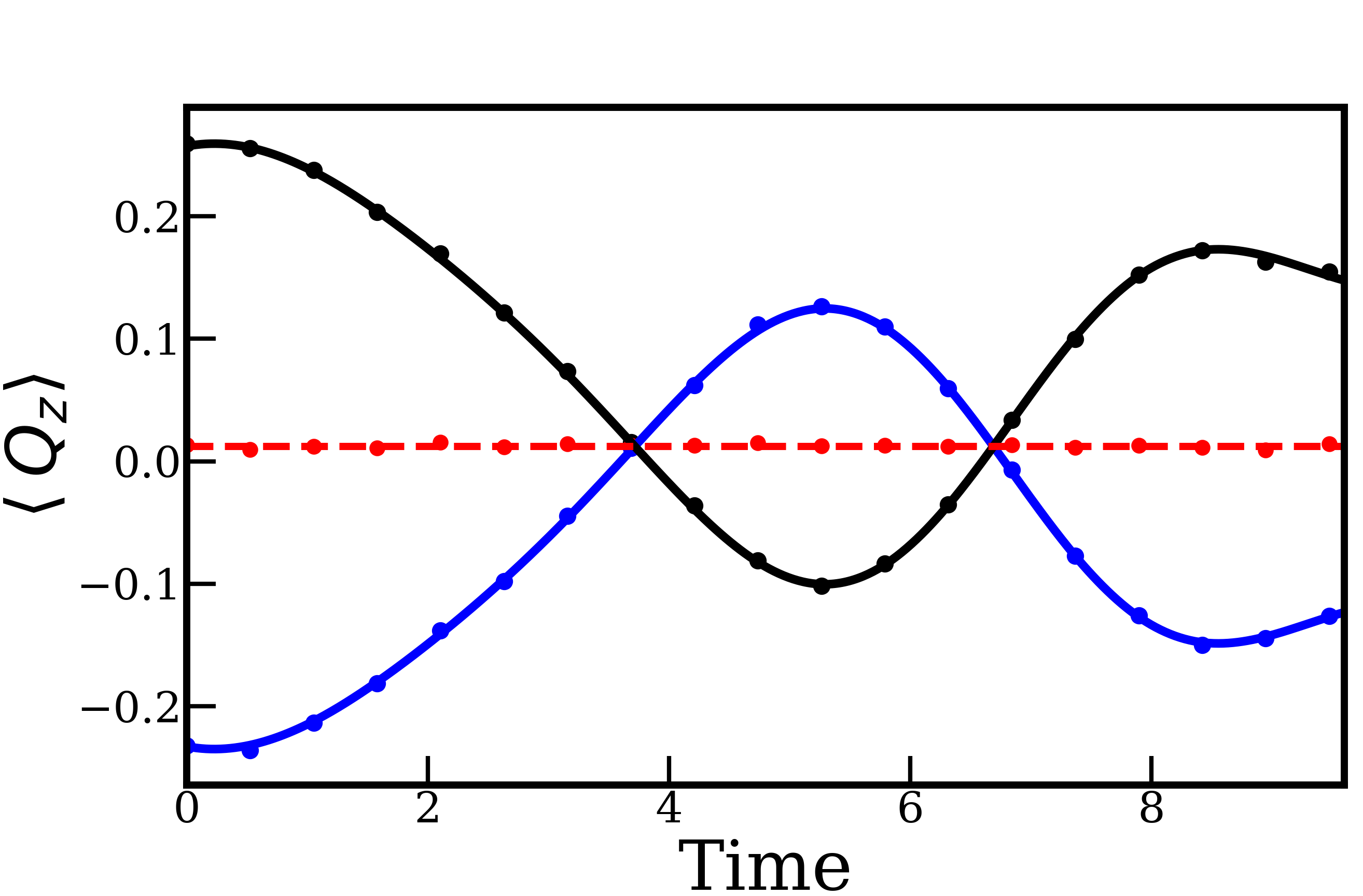}
    \includegraphics[height=130px]{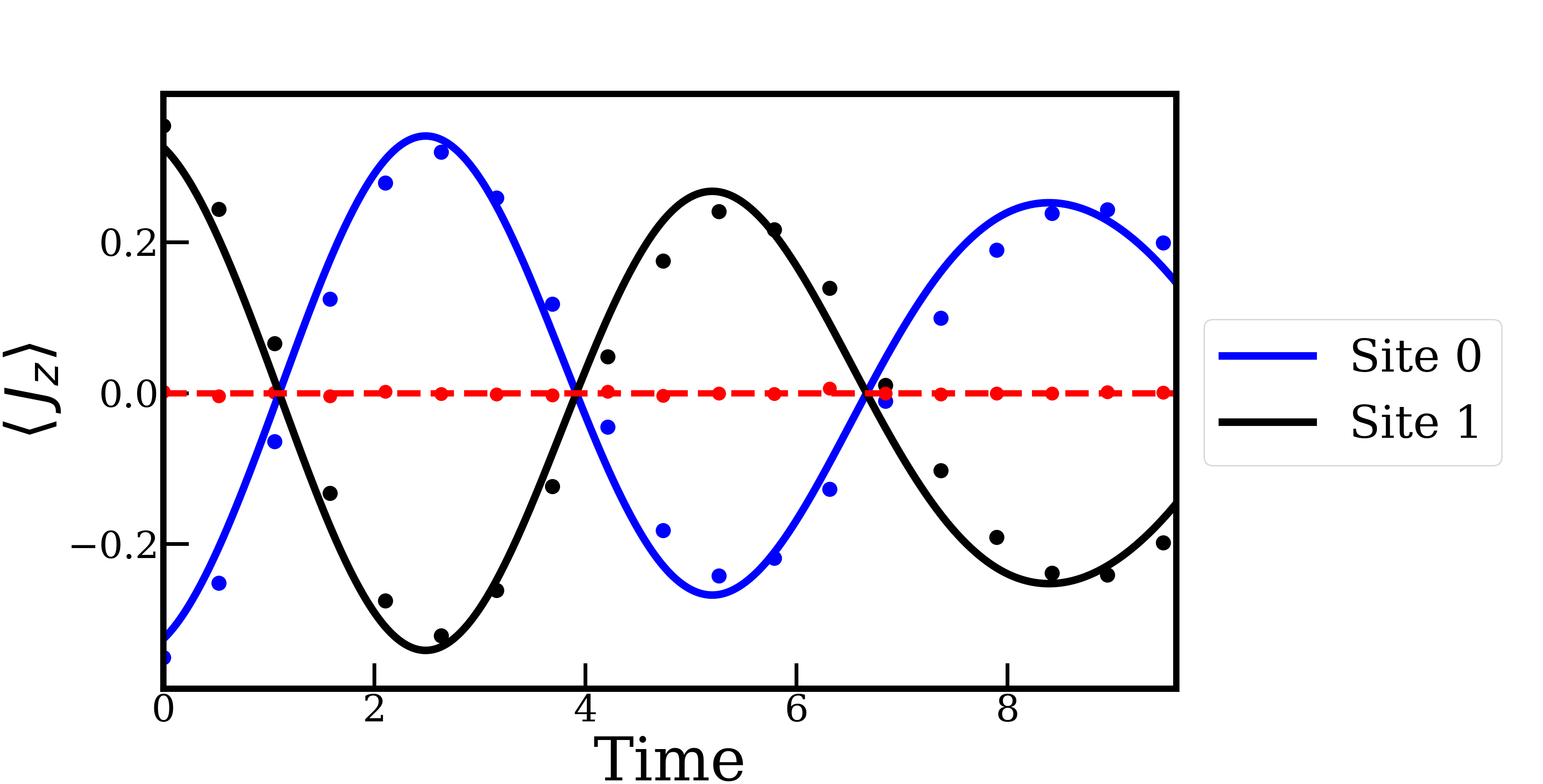}
    \caption{Time-dependence of the expectation value of the $z$-component of the $O(3)$ Noether charge (left) and current (right) predicted using shadow tomography for a two-site system, prepared in a random initial state, with $J_r = 0.1$. Solid lines indicate the exact charge or current on each site while points show the values predicted using shadow tomography. Red dashed line and dots show the exact and predicted values, respectively, of the total charge or current across both sites. This is expected to be a constant for the charge and zero for the current. Since $Q_z$ has range $k=1$, the results converge more rapidly than suggested by~\cref{eqn:HuangNoether}. Here we have used $N = 1,000,000$ and $5,000,000$ random Pauli measurements for the charge and current respectively with $n = 10$ per time-step ($\delta \approx 0.027 $). This implies $\varepsilon \approx 3.7\times 10^{-3}$ and ${}\approx 0.030$ for the charge and current respectively. Note that, even though we have used more measurements for the current, the prediction is less accurate than that for the charge owing to the four-fold increase in the range of the observable.}
    \label{fig:charge}
    \label{fig:current}
    \label{fig:measurement}
\end{figure*}

Concretely, we measure the time-dependence of local observables like the Noether current through several steps. 
\begin{enumerate}
    \item Prepare the desired initial state.
    \item Simulate time-evolution for time $t$.
    \item Perform a random unitary operation (this could be a random Clifford~\footnote{The Clifford group is useful for many concepts in quantum information. The n-qubit Clifford group $C_n$ is defined as the set of unitaries which normalize the $n$-qubit Pauli group $P_n$. That is, if $U \in U(2^n)$, then $U \in C_n$ if for every $V \in P_n$ there is a $W \in P_n$ such that $UVU^{\dagger} = W$.} circuit as in Ref.~\onlinecite{2002.08953}), then measure the state in the computational basis.
    \item Repeat steps \numrange[range-phrase = --]{1}{3} $N$ times. Use the results to predict, through a median-of-means estimator described below, the measurement outcome of the observable at time $t$ to precision $\varepsilon$. It is known that one requires $N = \mathcal{O}(\text{log}(\mathcal{V}/\varepsilon^2))$.
    \item Repeat steps \numrange[range-phrase = --]{1}{4}, sweeping $t$ over a range of values from $t_{\text{initial}}$ to $t_{\text{final}}$.
\end{enumerate}
For the present case we choose a random unitary ensemble composed of single-qubit Clifford circuits. This choice is equivalent to measuring the state in a random Pauli basis at each step. Then the results of Ref.~\onlinecite{2002.08953} tell us that to estimate the expectation value of a local observable $\mathcal{O}$ acting non-trivially on only $k$ qubits (at all sites on a lattice of volume $\mathcal{V}$) to within $\varepsilon$ of its true value with probability $1-\delta$, it suffices to repeat steps 1--3 $N$ times, where
\begin{equation}\label{eqn:Huang}
\begin{split}
    N &= (2\text{log}(2\mathcal{V} /\delta))\frac{34}{\varepsilon^2} 4^k||\mathcal{O}||^2_{\infty}.
\end{split}
\end{equation}
For $\mathcal{O} = J_{z,x}$ in one spatial dimension, we have
\begin{equation}\label{eqn:HuangNoether}
\begin{split}
    N = (2\text{log}(2\mathcal{V} /\delta))\frac{8704}{\varepsilon^2} |J_r|^2.
\end{split}
\end{equation}
For instance, estimating the current on a two-site system to two decimal places with success probability $90\%$ ($\delta = 0.1$, $\varepsilon = 0.01$, $k=4$) is guaranteed provided $N \gtrapprox (5\times 10^7) |J_r|^2$. Since each repetition can be performed in parallel, such large values of $N$ pose no technical difficulties.

The precision parameters $\varepsilon$ and $\delta$ are set independently via a simple median-of-means protocol. That is, we construct $N$ classical representations of the initial state using the procedure outlined above, split them into $n$ groups of $N/n$ shadows each, and average over each group. This gives a set of $n$ classical states $\{\rho_1,\cdots,\rho_n\}$. We return the median expectation value, $\text{Median}(\{ \text{Tr}(\mathcal{O}\rho_1),\cdots,\text{Tr}(\mathcal{O}\rho_n)\})$, of the desired observable over this set, yielding rigorous guarantees on $\delta$ and $\varepsilon$ as in Ref.~\onlinecite{2002.08953};
\begin{equation}\label{eqn:guarantee}
\begin{split}
    n &= 2\text{log}(2\mathcal{V}/\delta) \\
    N &= \frac{34 n}{\varepsilon^2}4^k|O|^2_{\infty}.
\end{split}
\end{equation}

As a demonstration of principle, we compute the time-evolution of the charge $Q_z$ and its current on a two-site system prepared in a random initial pure state, that is, one whose complex-valued entries are identically and independently distributed according to a normal distribution. The results are shown in~\cref{fig:measurement}. In both cases, the results appear significantly better than the rigorous performance guarantees suggested by \cref{eqn:guarantee}. The errors in the current are larger than those in the charge owing to the fact that the charge is a sum of single qubit operators, whereas the current acts on four qubits at a time. The closeness of the predicted values to the exact results suggest that this method may be an effective way to measure dynamic quantities in lattice models of quantum field theories on a quantum computer.

\section{Error Analysis}\label{sec:err}
% In this section, we analyze the various sources of error and give resource estimates for this model perturbatively in powers of $J_r$. Specifically, we seek to prepare the ground state $\ket{\Omega(J_r)}$ to within precision $\varepsilon$ starting from the zero-coupling ground state $\ket{\Omega(0)}$, where precision is any suitable measure of infidelity between the target state $\ket{\Omega(J_r)}$ and the output of the quantum algorithm.

Assuming perfect gate fidelity, there are two sources of errors we need to consider---Trotterization and violations of adiabaticity. Trotter error, which is $\mathcal{O}(\Delta t^2)$ in a first-order approximation, can be reduced at the cost of larger circuit depth by using a Trotter-Suzuki approximation of sufficiently high order, however we choose to work with a first-order approximation for several reasons. First, the algorithm of \cref{sec:nisq} was designed only to second order in the Trotter step size $\Delta t$, so any improvement via higher-order formulae would be lost.  Second, the field theory of interest emerges at a quantum critical point and depends primarily on the symmetries preserved by the Hamiltonian, rather than its exact form. Under reasonable assumptions, the errors induced by Trotterization reside in the algebra of operators invariant under the same symmetries as the original Hamiltonian. Thus, treating Trotterized time-evolution as exact evolution under an effective Hamiltonian by resumming the Baker-Campbell-Hausdorff formula, the effective Hamiltonian will likely reside in the same universality class as the original~\cite{inprep}. In that case, close to a critical point, the effective Hamiltonian and the target field theory differ only in a set of irrelevant UV operators which should not affect the physical content of the lattice theory.

We mention, in passing, that throughout this work we use the fidelity of the prepared quantum state as a measure of accuracy. This is, however, known to be too strong.  High fidelity implies that all observables are close to their desired values, but the continuum field theory cares only about products of local operators at separations scaling as the divergent correlation length.  In other words, the relevant measure of fidelity should be calculated after tracing out the ultraviolet degrees of freedom. We ignore this subtlety in this work.

% Here we consider two parameter regimes: the weak-coupling and quantum-critical limits. Our analysis of the weak-coupling regime is based on perturbative calculations of the spectral gap in orders of $J_r$ given in \cref{sec:perturbative}, while that for the quantum-critical regime relies on knowledge of the critical exponents for the target continuum limit. 

\subsection{\label{sec:WeakErrorAnalysis}Weak-Coupling Limit}
In this section we quantify errors due to Trotterization and violation of the adiabatic condition within the weak coupling approximation $J_r \ll 1$.

First, we will consider errors due to non-adiabaticity. Let $\widetilde{H}(t)$ be a time-varying Hamiltonian where $0 \leq t \leq T$, $H(\tau) = \widetilde{H}(t/T)$, and the energy gap of $H(\tau)$ is $\Delta(\tau)$. A theorem due to Teufel~\cite{Teufel} says that as long as the total time of the adiabatic evolution $T$ satisfies
\begin{equation}\label{eqn:teufel}
\begin{split}
    T \geq \frac{4}{\varepsilon}\scalebox{2.3}{[} &\frac{||\dot{H}(0)||}{\Delta(0)^2} + \frac{||\dot{H}(1)||}{\Delta(1)^2} \\ 
    &+ \int\limits_0^1 d\tau \left(10\frac{||\dot{H}(\tau)||^2}{\Delta(\tau)^3} + \frac{||\ddot{H}(\tau)||^2}{\Delta(\tau)^3} \right)\scalebox{2.3}{]},
\end{split}
\end{equation}
then the final state is within $\varepsilon$ of the true ground state. We can schedule the adiabatic evolution by linearly interpolating the coupling from its initial to its final value, using a series of discrete time steps $\Delta t_i$ where $\Delta t_i$ is set by the inverse spectral gap. According to our perturbative estimate of the spectral gap (\cref{sec:perturbative}),
\begin{equation}
    \Delta t_i \coloneqq 1/\Delta_i = 1/(1-2dJ_{r,i}+3dJ_{r,i}^2).
\end{equation}
The linear interpolation ensures that $\ddot H(\tau) = 0$. Since the on-site term is time-independent, it does not factor into the adiabatic condition. Then, defining $H_{h,\langle x, x' \rangle}$ as the hopping term $H_h$ from \cref{eqn:model} restricted to the lattice sites $x,x'$ (and similarly for $H_p$), we have
\begin{equation}
    \dot{H}(\tau) = J_{r,\text{max}} \sum_{\langle x,x'\rangle } (H_{p,\langle x, x' \rangle} + H_{h,\langle x, x' \rangle}),
\end{equation}
where \(J_{r,\text{max}}\) is the final desired value of \(J_r\) and
\begin{align}
%\begin{split}
    ||\dot{H}(\tau)|| &=& ||J_{r,\text{max}}\sum_{\langle x,x'\rangle} (H_{p,\langle x, x' \rangle} + H_{h,\langle x, x' \rangle})|| \nonumber\\ &\leq& J_{r,\text{max}}\sum_{\langle x,x'\rangle}||H_{p,\langle x, x' \rangle} + H_{h,\langle x, x' \rangle}|| \nonumber\\
    &\propto& \mathcal{V}J_{r,\text{max}}.
%\end{split}
\end{align}

In this analysis, we ignored the fact that we change the Hamiltonian along a staircase approximation to the linear function. Assuming that $\Delta(\tau)$ decreases monotonically with $\tau$, a standard result from real analysis lets us bound the error induced by replacing the integral in \cref{eqn:teufel} with a sum as
\begin{equation}
\begin{split}
    10&\Delta t_{\text{max}} \left(\frac{||\dot{H}(0)||^2}{\Delta(0)^3} - \frac{||\dot{H}(1)||^2}{\Delta(1)^3}\right)\\
    & = \mathcal{O}(\mathcal{V}^2 J_{r,\text{max}}^3\Delta t_{\text{max}})
    \end{split}
\end{equation}
where we have defined $\Delta t_{\text{max}} = \text{max}_i\{\Delta t_i\}$. To second order in $J_{r,\text{max}}$, we can obtain the mass gap perturbatively and choose
\begin{equation}
    \Delta t_i = \left(1-2d\left(\frac{iJ_{r,\text{max}}}{N}\right) + 3d\left(\frac{iJ_{r,\text{max}}}{N}\right)^2\right)^{-1}\,.
\end{equation}
Since for small $J_{r,\text{max}}$ the spectrum is completely gapped, $\Delta t_i$ is bounded by a constant. Therefore, the leading discretization error is of order $\mathcal{V}^2J_{r,\text{max}}^3$. 

To summarize, to second order in $J_{\max}$, the adiabatic condition becomes
\begin{align}
    T &= \mathcal{O}\left(\frac{J_{r,\text{max}}^2\mathcal{V}^2}{\varepsilon}+\mathcal{V}^2J_{r,\text{max}}^3\right) \\
    &= \mathcal{O} \left(\frac{J_{r,\text{max}}^2\mathcal{V}^2}{\varepsilon}\right).
\end{align}
This result holds so long as the energy gap is bounded from below by a constant. For fault-tolerant devices, the limiting factor is then the adiabatic evolution near a phase transition where the gap shrinks to zero. In that case, knowledge of the critical exponent can determine the optimal scheduling as in Ref.~\onlinecite{Jordan1130}. 

Having obtained an upper bound on the error due to non-adiabaticity, we now include that arising from Trotterization. Suppose there are $N$ time steps and the coupling is linearly interpolated between $0$ and $J_{r,\text{max}}$. Since each time-evolution operator has leading error corrections of order $\mathcal{V} \Delta t_i^2$ in a first-order Trotter approximation, the full evolution is valid to order $\mathcal{V} \sum_i \Delta t_i^2$. Naturally, this sum is bounded by $\mathcal{V} T^2$, so demanding that the Trotterization error is of order $\varepsilon$ constrains the total time simulated as
\begin{equation}
    T = \mathcal{O}\left( \sqrt{\frac{\varepsilon}{\mathcal{V}}}\right).
\end{equation}
Na\"\i vely, we assign as the total error the sum of those induced by Trotterization and non-adiabaticity, so that the total error is of order $\varepsilon$ provided that
\begin{equation}
    T = \mathcal{O}\left(\frac{(J_{r,\text{max}}\mathcal{V})^2}{\varepsilon} + \sqrt{\frac{\varepsilon}{\mathcal{V}}}\right).
\end{equation}

The total run-time is
\begin{equation}
    T = \sum \Delta t_i = \mathcal{O}(N),
\end{equation}
so equivalently we can write 
\begin{equation}
    N = \mathcal{O}\left(\frac{(J_{r,\text{max}}\mathcal{V})^2}{\varepsilon} + \sqrt{\frac{\varepsilon}{\mathcal{V}}}\right).
\end{equation}
Previously we showed that a single time-step can be implemented exactly with a constant number of gates per lattice site, so, multiplying by the total number of links, the total time-complexity of our algorithm is
\begin{equation}
    \mathcal{O}\left(\frac{J_{r,\text{max}}^2\mathcal{V}^3}{\varepsilon} + \sqrt{\varepsilon \mathcal{V}}\right)
\end{equation}
in arbitrary spatial dimension. We will continue to display the subdominant term in this expression to keep track of the Trotterization error in comparison to the adiabaticity error.

\subsection{Quantum-Critical Regime}
The estimates in the previous section describe the resource requirements for preparing the ground-state of the discretized theory in the weak-coupling limit. Our goal, however, is to probe the physics of a quantum field theory, the $O(3)$ sigma model, which emerges only in the long-distance limit as the correlation length in the theory diverges. In two and three spatial dimensions the model in \cref{eqn:model} resides in the universality class of the $O(3)$ non-linear sigma model, meaning that, in infinite volumes, it undergoes a quantum phase transition at some critical value of the coupling $J_{r,c}$. Accordingly, the mass gap $\Delta (J_r)$ scales as $|J_r-J_{r,c}|^{\nu}$, where $J_{r,c}$ and $\nu$ are the critical coupling and exponent, respectively. For our model, we have efficient classical procedures for determining the critical parameters~\cite{PhysRevD.100.054505}: $\nu = 0.693(15)$ in (2+1)-dimensions and $\nu = 0.5050(96)$ in (3+1)-dimensions. The critical values of the coupling are $J_{r,c} = 4.81695(37)$ in (2+1)-dimensions and $J_{r,c} = 10.09817(55)$ in (3+1)-dimensions.

To estimate the time required to prepare the ground state at coupling $J_r$ close to $J_{r,c}$, we apply the same methods as in \cref{sec:WeakErrorAnalysis}, starting from \cref{eqn:teufel}. We find that the error due to non-adiabaticity is less than $\varepsilon$ provided that the total simulated time obeys
\begin{equation}
  \begin{split}
    T \geq {}&\frac{4\mathcal{V}||J_r (H_p + H_h)||}{\varepsilon}\big{(}1 + \alpha |J_r -J_{r,c}|^{-2\nu}\\
                                               &\qquad{}+10\mathcal{V}||J_r (H_p + H_h)||\int\limits_{0}^{1} ds \Delta(s)^{-3}\big{)},
  \end{split}
\end{equation}
where \mbox{$\Delta(\tau) \sim \alpha|J_r-J_{r,c}|^{\nu}$} for $\tau \approx 1$, the linearly interpolated Hamiltonian is
\begin{equation}
    H(s) = \sum\limits_x H_{1,x} + J_r s\sum\limits_{\langle x,x'\rangle }(H_{p,\langle x,x'\rangle } + H_{h,\langle x,x'\rangle})
\end{equation}
and $s$ ranges from $0$ to $1$. 

Because the gap approaches zero as $J_r$ approaches $J_{r,c}$, we assume that $\Delta(s)$ is bounded from below by $\Delta(1)$. Since the norm of the nearest-neighbor Hamiltonian $||(H_{p,\langle x, x' \rangle} + H_{h,\langle x, x' \rangle})||$ on a single pair of sites is of unit order, we can lower bound the total required simulated time $T$ in our algorithm as
\begin{equation}
    T \geq \frac{4\mathcal{V}|J_r|}{\varepsilon}\big{(}1+\alpha|J_r-J_{r,c}|^{-2\nu}+10\mathcal{V}|J_r||J_r-J_{r,c}|^{-3\nu}\big{)}.
\end{equation}
For $J_r$ close to the critical value, then, the dominant contribution is
\begin{equation}
    T = \mathcal{O}\left(\frac{|J_r \mathcal{V}|^2}{\varepsilon |J_r -J_{r,c}|^{3\nu}}\right).
\end{equation} 
Trotterization again requires that $T = \mathcal{O}(\sqrt{\varepsilon/\mathcal{V}})$, so the total time-complexity of adiabatic state preparation in the quantum-critical regime in arbitrary spatial dimension is
\begin{equation}
    \mathcal{O}\left(\frac{|J_r|^2\mathcal{V}^3}{\varepsilon |J_r-J_{r,c}|^{3\nu}}+\sqrt{\varepsilon \mathcal{V}}\right).
\end{equation}
In particular, this result implies that we can prepare the ground state of our model at coupling $J_r$ close to the critical coupling $J_{r,c}$ in time polynomial in the distance to the critical point, lattice volume, and precision $\varepsilon$. The scaling of the time-complexity with the critical exponent $\nu$ follows directly from the adiabatic condition in \cref{eqn:teufel}, and is independent of the spatial dimension. For infinite lattice size and $J_r$ sufficiently close to $J_{r,c}$, physical properties of this ground state correspond to those of the vacuum state of the non-linear $O(3)$ sigma model.

\section{\label{sec:nisq}Quantum Compiling and Randomization}

Current quantum devices are severely limited by noisy gate implementations and short decoherence times. As a result, real near-term simulations demand circuits which can be implemented in time on the order of the qubit lifetime, strongly favoring short-depth circuits. Since, as in \cref{sec:CI}, implementing the time-evolution operator exactly using the circuit we provided would require ancilla qubits and at least $58$ gates per time-step, realizing the time-evolution operator of our model does not appear to be feasible for a near term device. Therefore, we should consider alternative approximate methods.

In order to find a short-depth circuit approximating the time-evolution operator for the nearest-neighbor Hamiltonian in \cref{eqn:model}, we use a classical simulation of quantum-assisted quantum compiling (QAQC), originally proposed by Khatri et al.~\cite{Khatri2019quantumassisted}. Just as a compiler translates high-level source code to a low-level version which can be interpreted by a computer processor, quantum compiling seeks to translate a representation of a unitary, for instance, its matrix representation in a fixed basis, to a sequence of gates executable on a quantum device. \looseness-1

We fix an allowed set of gates specific to a device and set a desired circuit depth, then apply standard optimization techniques to obtain an approximation of the desired quantum circuit. Here, we allow only CNOT and arbitrary single-qubit gates, as in IBM's QX architecture. The cost function is taken to be the square of the Frobenius norm of the difference between the exact operator and its approximation.

\begin{figure}[t]
    \centering
    \includegraphics[width=230px]{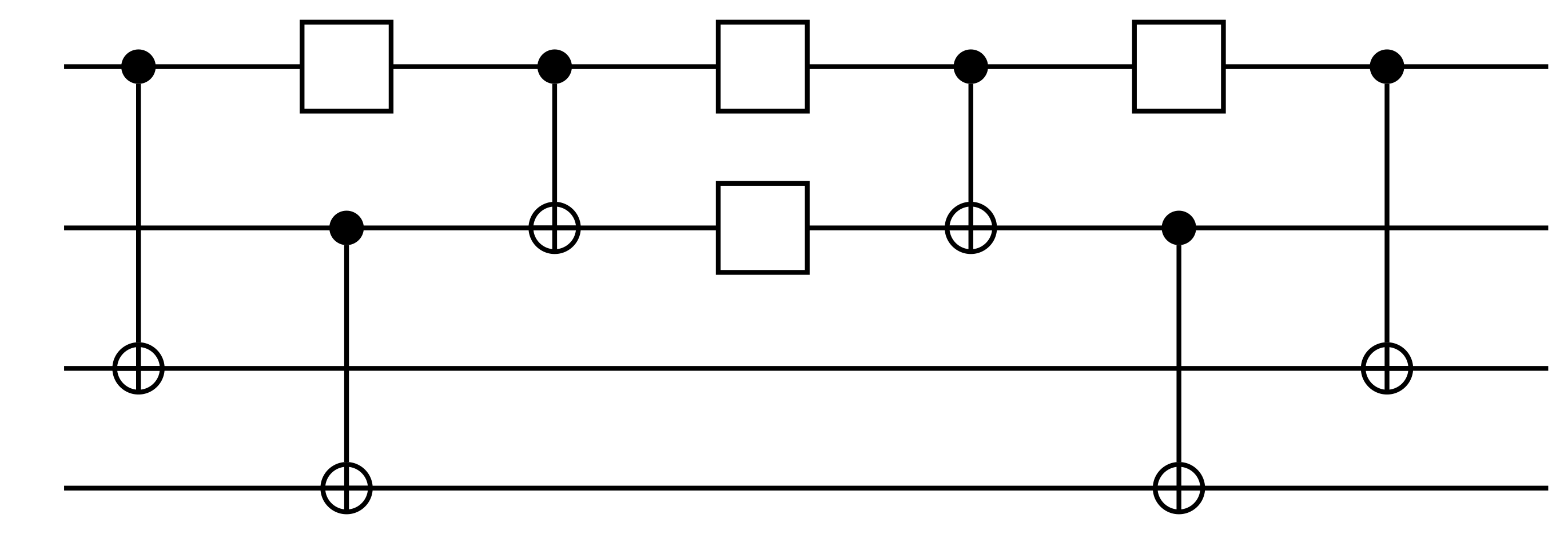}
    \caption{Circuit for implementing nearest-neighbor time evolution step using ten gates. The single qubit operations are optimized over all possible unitary gates.}
    \label{fig:my_label3}
\end{figure}
Performing this protocol for our model, we obtain 10, 20, and 30-gate circuits composed only of CNOTs and single-qubit gates (see \cref{fig:my_label3}) approximating the time evolution operator on a single pair of sites,
\begin{align}
    U(J_r \Delta t) = e^{i J_r \Delta t (H_h + H_p)},
\end{align}
for a fixed value of the coupling constant and time step ($J_r = 0.04$, $\Delta t = 0.2$). In each case, the optimizer was allowed to run for up to 48 CPU hours. One could repeat this procedure for each value of the coupling needed, however, for simulations involving many steps, this would be computationally demanding.

Instead, we run the procedure a single time and interpolate with a randomized approach~\cite{randomized}. By performing a probabilistically scheduled sequence of gates, the dynamics of the quantum state is represented as a non-unitary super-operator, also known as a quantum channel, mapping density matrices to density matrices. The aim is to produce a mixed state that still has a very high overlap with the desired ground-state, so that observables and expectation values are reproduced with little error.

\begin{figure}[t]
    \centering
    \includegraphics[width=230px]{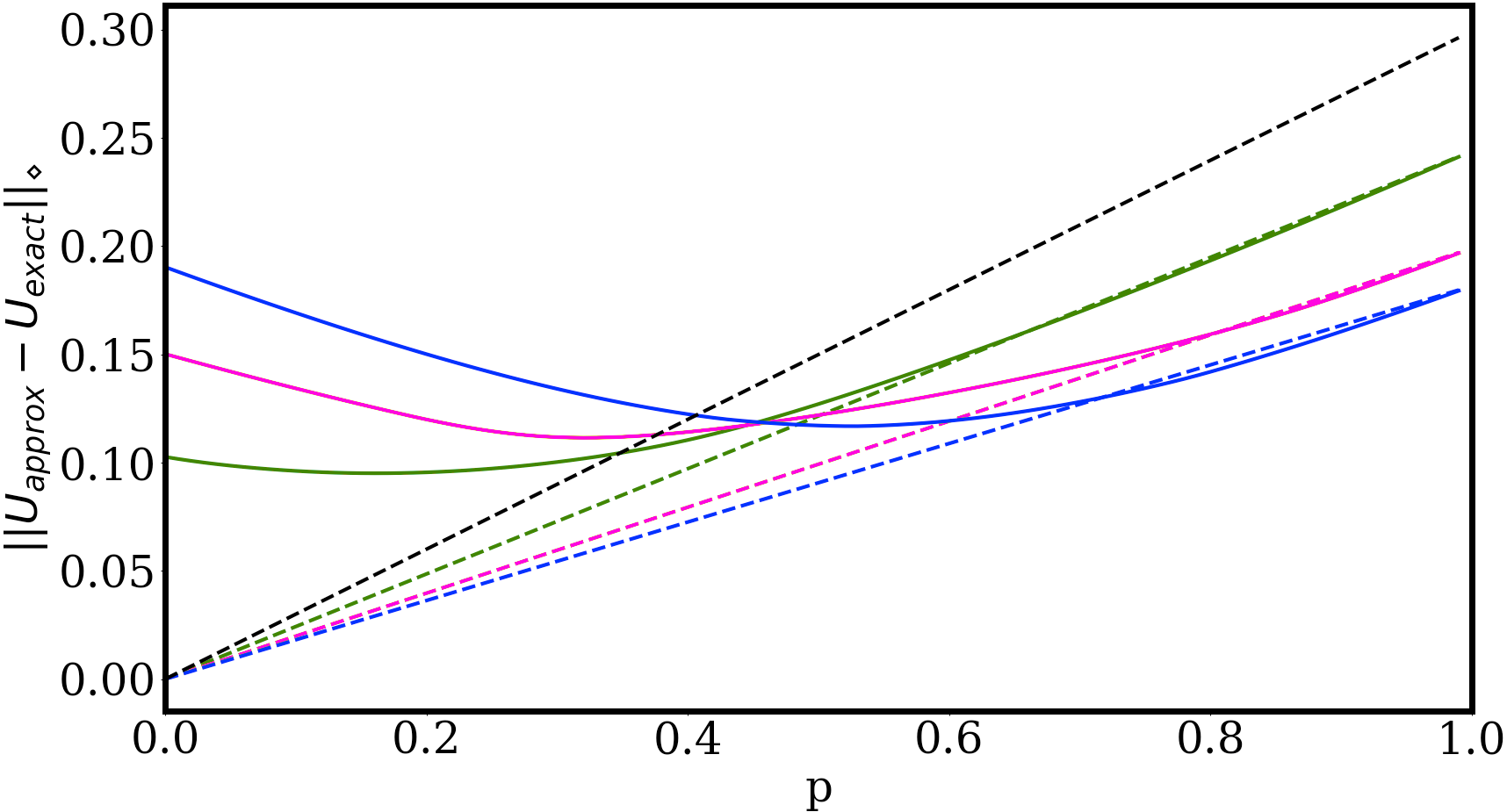}
    \caption{Diamond distance, given by \cref{eqn:diamond}, between the exact time-evolution operator $e^{iJ_r \Delta t(J_r) (H_p+H_h)}$ and its approximation by short-depth unitaries, on four qubits as a function of $p = \frac{J_r \Delta t(J_r)}{\max(J_r \Delta t(J_r))}$, shown here for $\max (J_r\Delta t(J_r)) = 0.1$. Curves are shown for 10 (blue), 20 (magenta), and 30 (green) gate circuits. Solid lines indicate deterministically applying the approximate circuit at each time-step. Dashed lines correspond to applying the identity with probability $1-p$ and the approximate circuit with probability $p$. The black dashed line shows the identity circuit as a benchmark for performance. Randomized approach appears to work better than deterministic one for all circuit depths, since the randomized circuit is closer to the exact time-evolution operator in diamond distance.}
    \label{fig:diamond}
\end{figure}

\begin{figure}[t]
    \centering
    \includegraphics[width=230px]{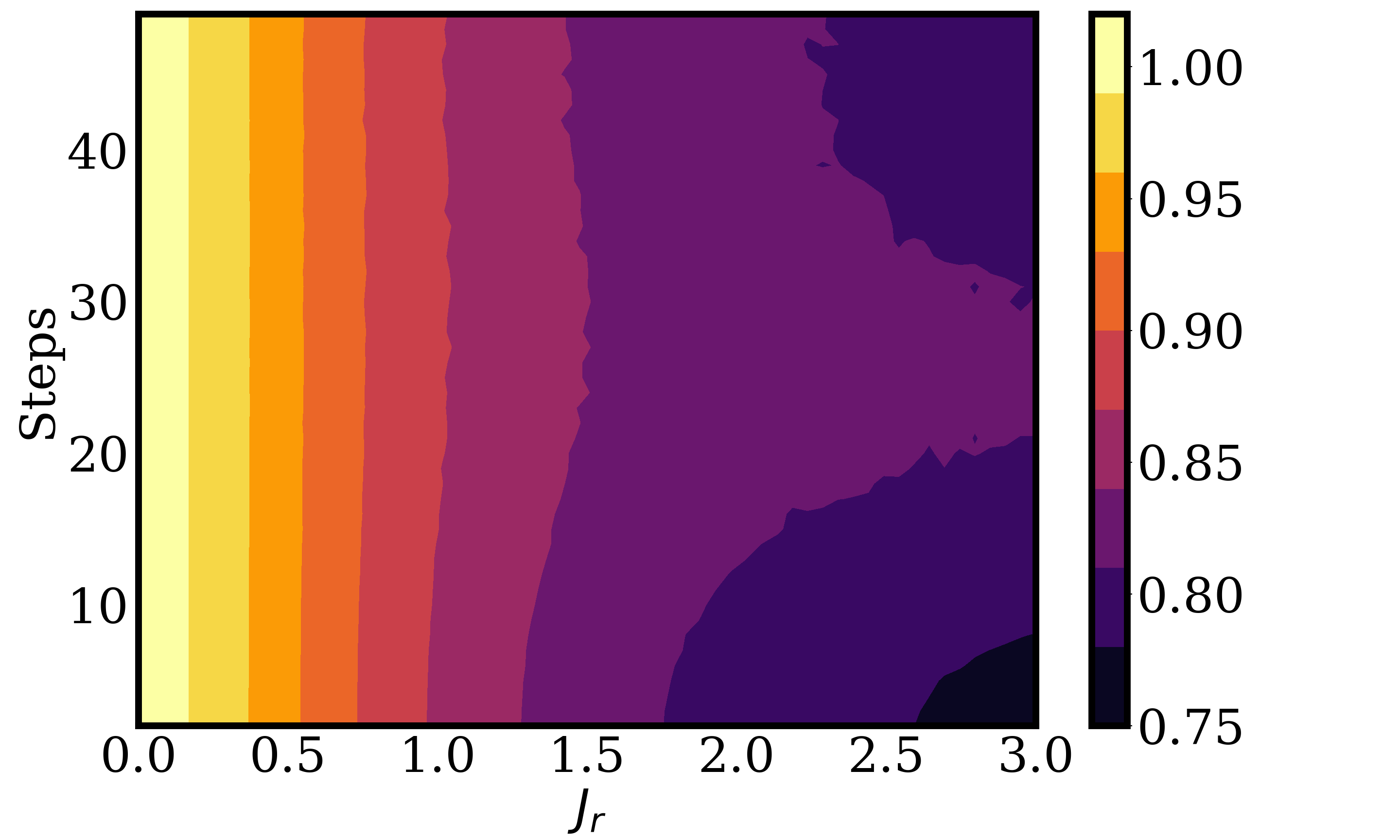}
    \caption{Fidelity of adiabatic state preparation on four qubits for several numbers of Trotter steps using the short-depth probabilistic method with circuit depth 10. Fidelity is averaged over 10 repetitions. There appears to be only slight variation in the accuracy as the number of steps is increased. Note that the fidelity does not increase monotonically with the number of steps. This is because the error from quantum circuit compiling grows additively with the number of steps, eventually out-pacing whatever gains are made by reducing Trotter error through more steps.}
    \label{fig:adiabaticstateprep}
\end{figure}

\begin{figure}[t]
    \centering
    \includegraphics[width=230px]{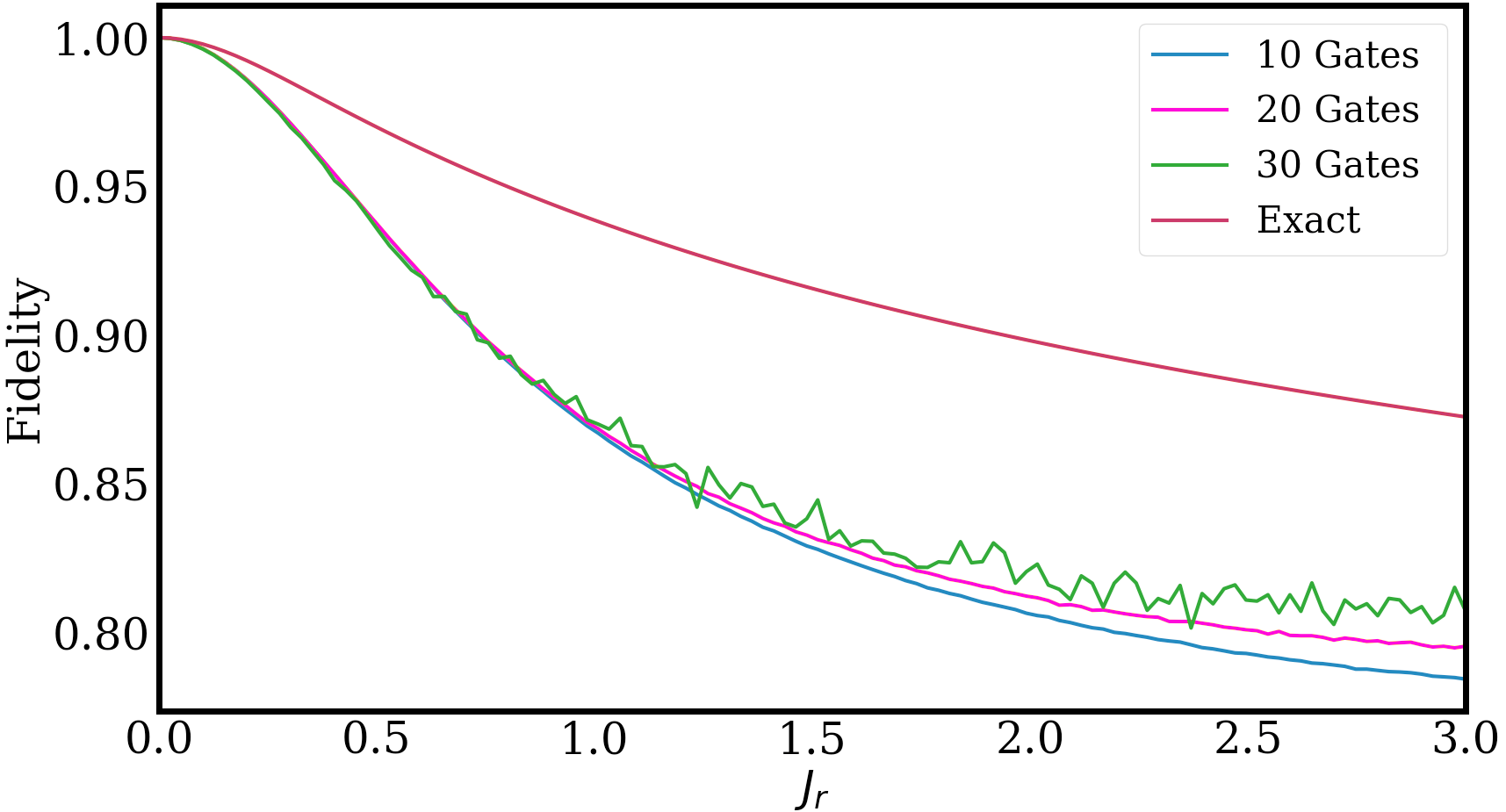}
    \caption{Adiabatic state preparation on four qubits using probabilistic method with 10, 20, and 30-gate circuits, and exact nearest-neighbor time-evolution operator. For the curve labeled 'exact,' deviations from one are due entirely to Trotterization and non-adiabatic errors. Results are averaged over 100 repetitions, using 10 Trotter steps. Results for short-depth circuits are slightly worse than an exact Trotterized approach, with small improvements through larger depth. We expect the short-depth circuits to perform better in the presence of noise.}
    \label{fig:gatedepth}
\end{figure}

The specific random sequence of gates is determined as follows. First we determine $U_{\text{approx}}$ as the short-depth circuit best approximating \(U(J_r \Delta t)\) for the largest \(J_r\Delta t\) of interest. Then, for adiabatic state preparation with $N$ steps, and coupling $J_{r,i}$ and time step $\Delta t(J_{r,i})$ at the $i$'th step, let 
\begin{align}
p_i = \frac{J_{r,i} \Delta t(J_{r,i})}{\max_i(J_{r,i}\Delta t (J_{r,i}))}.
\end{align}
At each time step, we apply the identity circuit with probability $1-p_i$ and  $U_{\text{approx}}$ with probability $p_i$. This has the advantage of optimality for fixed circuit depth at $p_i = 0, 1$. Since the randomized circuit is \textit{not} a unitary operation, one needs to modify our measure of fidelity. We choose to use the channel fidelity, namely the diamond norm of the difference between the approximate and exact circuits. The diamond norm is a measure of how hard it is to distinguish two quantum channels with a single measurement. Specifically, if $\Phi, \Phi' \in \mathcal{L}(H\otimes H) \rightarrow \mathcal{L}(H\otimes H)$ are two unital quantum channels, their diamond distance is
\begin{equation}\label{eqn:diamond}
    ||\Phi-\Phi'||_{\diamond} = \max_{\rho} \text{Tr}(\Phi \otimes I)(\rho) - (\Phi' \otimes I)(\rho)),
\end{equation}
where $\rho$ is a density matrix on the k-qubit Hilbert space operated on by the circuit, with the addition of $k$ possibly entangled ancilla qubits not acted on by the circuit. We use semidefinite programming to perform the maximization in \cref{eqn:diamond} for each circuit. The results are shown in \cref{fig:diamond},~\cref{fig:adiabaticstateprep} and~\cref{fig:gatedepth}.

\subsection{Analysis of randomized algorithm}
Consider an arbitrary Hamiltonian $H$ which is a function of a set of coupling constants $\lambda_i$. Suppose we can implement $U(\lambda_i,\Delta t) = e^{-iH(\lambda_i)\Delta t }$ exactly for two sets of values for the couplings and time step; $\lambda_{i}(0), \Delta t(0)$ and $\lambda_{i}(1), \Delta t(1)$. In order to find the ground state of $H(\lambda_{i}(1))$ using adiabatic state preparation (or similar quantum quench), one could prepare the system in the ground state of $H(\lambda_{i}(0))$ and linearly interpolate the couplings between $\lambda_{i}(0)$ and $\lambda_{i}(1)$. 

Now we form the quantum channel
\begin{equation}\label{eqn:channel}
    \Lambda_t (\rho) = (1-p)U(0)\rho U^{\dagger}(0) + pU(1)\rho U^{\dagger}(1),
\end{equation}
where $U(t)$ is shorthand for $U(\lambda_i(t),\Delta t(t))$. Letting $\widetilde{\lambda}_i(t) = \lambda_i(t)\Delta t(t)$ and $\lambda$ be any of the various couplings, we set for the probability
\begin{equation}
p(t) = \frac{\widetilde{\lambda}(t)-\widetilde{\lambda}(0)}{\widetilde{\lambda}(1)-\widetilde{\lambda}(0)}.
\end{equation}
Expanding \cref{eqn:channel} in powers of $\Delta t(t)$, we obtain
\begin{equation}
    \Lambda_t(\rho) = U(t)\rho U^{\dagger}(t)(1+\mathcal{O}(\Delta t^2)).
\end{equation}
Thus, we see that the time-dependent dynamics is reproduced to the same order in $\Delta t$ as in first-order Trotterization, implying that the randomized scheme does not change the asymptotic scaling of the time-complexity of our algorithm.

If, as in the present case, we replace the exact implementation of $U(1)$ with an approximate unitary $\overline{U}(1) = U(1)(1+\mathcal{O}(\delta))$, the result becomes
\begin{equation}
    \Lambda_t(\rho) = U(t)\rho U^{\dagger}(t)(1+\mathcal{O}(\Delta t^2,\delta)).
\end{equation}
In this way, it suffices to compile a constant number of circuits and interpolate between them using randomized simulation, rather than compiling a number of circuits growing linearly with the number of time steps.

\section{\label{sec:conc}Conclusion}
Quantum computing has enormous potential for future investigations in high energy physics. In this paper, we focused on the task of simulating a qubit-regularized version of the non-linear $O(3)$ sigma model, in particular, preparing its ground state and measuring dynamic quantities. We found that, in $d$-dimensions, for a lattice of size $\mathcal{V} = dL^d$, the ground state may be prepared near the quantum critical point to precision $\varepsilon$ in time $\mathcal{O}\left(\frac{\mathcal{V}^3|J|^2}{\varepsilon |J-J_c|^{3\nu}}+\sqrt{\varepsilon \mathcal{V}}\right)$ using ordinary Trotter methods and provided an explicit circuit representation. We described how to use shadow tomography to measure the time-dependent $O(3)$ Noether current efficiently on a near-term device to precision $\delta$ in time $\mathcal{O}(\text{log}(\mathcal{V}/\delta^2))$. Lastly, we performed numerical experiments simulating a heuristic algorithm for obtaining short-depth circuits approximating adiabatic ground state preparation for applications on intermediate-term devices, as well as an improved version implementing techniques from randomized quantum simulation.
\section*{Acknowledgment}
We would like to thank R. Somma, S. Chandrasekharan, H. Singh and J. Preskill for helpful discussions. TB and RG were funded under Department of Energy (DOE) Office of Science High Energy Physics Contract \#89233218CNA000001. LC was supported by the Laboratory Directed Research and Development program of Los Alamos National Laboratory under project number 20190065DR. LC was also supported by the U.S. Department of Energy (DOE), Office of Science, Office of Advanced Scientific Computing Research, under the Quantum Computing Application Teams program. AJB also acknowledges support from the Los Alamos Quantum Computing Summer School (QCSS) and its organizers. % \textcolor{red}{Need to acknowledge sponsors.}
\bibliographystyle{h-physrev-doi}
\bibliography{bibliography}
%\clearpage
%\newpage
%\mbox{~}
\onecolumngrid
  \hbox to\hsize{\hfill}
  \hbox to\hsize{\hskip0.23\hsize\leaders\hrule height 0.5pt depth 0.5pt\hfil\hskip0.23\hsize}
  \vspace*{-\baselineskip}
  \hbox to\hsize{\hskip0.3\hsize\leaders\hrule height 0.6pt depth 0.6pt\hfil\hskip0.3\hsize}
  \vspace*{-\baselineskip}
  \hbox to\hsize{\hskip0.37\hsize\leaders\hrule height 0.7pt depth 0.7pt\hfil\hskip0.37\hsize}
  \vspace*{-\baselineskip}
  \hbox to\hsize{\hskip0.43\hsize\leaders\hrule height 0.8pt depth 0.8pt\hfil\hskip0.43\hsize}
  \vspace*{2\baselineskip}
\twocolumngrid
\appendix

\section{Representation Theory}
\label{sec:reptheory}
The Hamiltonian for the lattice-regulated $O(3)$ non-linear sigma model is essentially that of an $O(3)$ rotor model;
\begin{equation}
    \hat{H} = J_1 \sum_{i} \frac{\hat{\vec{L}}_i^2}{2} - J_2\sum_{\langle i,j \rangle} \hat{\vec{\phi}}_i \cdot \hat{\vec{\phi}}_j := \hat{H}_{L}+\hat{H}_{\phi},
\end{equation}
where $i$ and $j$ are nearest neighbor sites on a Euclidean spatial lattice in d-dimensions, $\hat{\vec{\phi}}_i$ is a unit 3-vector associated to site $i$ and $\hat{\vec{L}}_i$ is the angular momentum. In spherical coordinates the dot product of two unit Euclidean 3-vectors can be expressed as 
\begin{equation}
\begin{split}
    \vec{\phi}_1 \cdot \vec{\phi}_2 = \sin(\theta_1)&\sin(\theta_2)\cos(\phi_1 - \phi_2)\\ +& \cos(\theta_1)\cos(\theta_2).
\end{split}
\end{equation}
According to the Peter-Weyl theorem, the space of square integrable functions on the unit sphere is isomorphic to the direct sum of all unitary irreducible representations of $SO(3)$. This is precisely the content of spherical harmonic analysis. In terms of spherical harmonics $Y^m_l(\theta,\phi)$,
\begin{eqnarray}
%\begin{split}
    \vec{\phi}_i \cdot \vec{\phi}_j &=& \frac{4\pi}{3}( Y^{0}_1(\theta_i,\phi_i)Y^0_1(\theta_j,\phi_j)\nonumber\\ 
  &&\qquad{}-Y^{1}_1(\theta_i,\phi_i)Y^{-1}_1(\theta_j,\phi_j) \nonumber\\ 
  &&\qquad{}-Y^{-1}_1(\theta_i,\phi_i)Y^1_1(\theta_j,\phi_j)).
%\end{split}
\end{eqnarray}
These spherical harmonics may be recast as Hermitian operators acting on a Hilbert space where the states are labelled by the irreps of $SO(3)$, as in Ref.~\onlinecite{PhysRevA.73.022328}. We define 
\begin{equation}
    \hat{Y}^m_l \ket{s} = \sqrt{2l + 1}\ket{l,m},
\end{equation}
where $\ket{s}$ is the $l = m = 0$ state, and the irrep states satisfy
\begin{equation}
    \braket{l,m | l',m'} = \delta(l,l') \delta(m,m').
\end{equation}
The remaining matrix elements of the $\hat{Y}^m_l$ are determined by Clebsch-Gordan decomposition
\begin{equation}
    \hat{Y}^{m_1}_{l_1} \hat{Y}^{m_2}_{l_2} = \sum_{L = |l_1-l_2|}^{L = l_1+l_2}\hat{Y}^M_L \braket{L,M|l_1,m_1;l_2,m_2},
\end{equation}
with $M = m + m'$.

We now imagine truncating the Hilbert space by imposing a hard cutoff on the irrep labels $L \leq L_{max}$. Setting $L_{max} = 1$ for the Hamiltonian of the non-linear O(3) sigma model, we have $4$ states per lattice site and a 16$\times$16 Hamiltonian;
\begin{equation}
    \hat{H}_{\phi} = \frac{4\pi J_2}{3} \sum_{\langle i,j \rangle, m} (-1)^m\hat{Y^{m}_{1}}(i)\hat{Y}^{-m}_{1}(j).
\end{equation}
Allowing this to act on the singlet-singlet state, we can identify a pair-creation term
\begin{equation}
    \hat{H}_p = \frac{4\pi J_2}{\sqrt{3}} \sum_{\langle i,j \rangle} (-1)^m \ket{m,-m}\bra{ss}_{i,j} + \text{h.c.}
\end{equation}
The Hamiltonian must either increase or decrease $l_1$ and $l_2$ by one, so the only other states on which $\hat{H}_{\phi}$ acts non-trivially are those of the form $\ket{sm}$ or $\ket{ms}$. Subtracting $\hat{H}_p$ from $\hat{H}_{\phi}$ yields a simple hopping term
\begin{equation}
    \hat{H}_h = \frac{4\pi J_2}{\sqrt{3}} \sum_{\langle i,j \rangle} \ket{sm}\bra{ms}_{i,j} + \text{h.c.},
\end{equation}
where
\begin{equation}
    \hat{H}_{\phi} = \hat{H}_h + \hat{H}_p.
\end{equation}
These results also follow by directly evaluating the integral
\begin{equation}
\begin{split}
    \int d\{\theta,\phi\}\  (\vec{\phi}_i \cdot \vec{\phi}_j)&\overline{Y}^{m_1}_{l_1}(\theta_1,\phi_1)\overline{Y}^{m_2}_{l_2}(\theta_2,\phi_2) \\ \times&Y^{m_3}_{l_3}(\theta_1,\phi_1)Y^{m_4}_{l_4}(\theta_2,\phi_2).
\end{split}
\end{equation}
Performing this calculation in Mathematica for all $\{l,m\}$, $l \leq 1$, we obtain precisely the terms above (up to normalization).

The kinetic term, $\hat{H}_{L}$, is diagonal in the $\ket{l_1,m_1;l_2,m_2}$ basis;
\begin{equation}
    \hat{H}_{L}\ket{l_1,m_1;l_2,m_2} = J_1\frac{l_1^2+l_2^2}{2}\ket{l_1,m_1;l_2,m_2},
\end{equation}
yielding an onsite potential term in the truncated Hilbert space,
\begin{equation}
    \hat{H}_{L} = \frac{J_1}{2}\sum_{m,i}  \ket{m}\bra{m}_i.
\end{equation}
In this way, the qubit model from \cref{eqn:model} can be interpreted as a truncation of the lattice-regulated Hamiltonian for the $O(3)$ non-linear sigma model. This provides a connection between our work and that of other groups who consider direct truncation schemes for simulating quantum field theories, as in Ref. \onlinecite{PhysRevLett.123.090501,PhysRevD.91.054506,PhysRevA.73.022328}.

\section{Perturbation Theory}
\label{sec:perturbative}
We consider the model of \cref{eqn:model} on a regular square or cubic lattice of length $L$ in $d$ dimensions. We wish to know the energy gap near the weak-coupling limit, since this will inform our adiabatic algorithm, and the fidelity between the strong-coupling ground state and that at finite coupling. Both questions may be handled by ordinary Rayleigh-Schr\"odinger perturbation theory. 

The weak-coupling vacuum $\ket{\Omega(0)}$ has the singlet on all spatial sites, $\ket{\Omega(0)} = \otimes_r \ket{s}_r$. In the degenerate case $\mu = 0$ at weak coupling, the first excited state manifold is spanned by all states with a single site in a triplet state. Thus, the energy gap at weak coupling is $1$ in units where $J = 1$. The first order correction to the ground state energy is zero. 

Since the first excited state manifold for a lattice with finite volume is highly degenerate, we must apply degenerate perturbation theory. Luckily, within this manifold the perturbation is easily diagonalized. Since the pair-creation term does not act within this subspace we may ignore it. The hopping term conserves $M$ and is translation invariant, so its eigenstates are also momentum eigenstates. Thus the hopping term decomposes into three equal blocks, each of which is diagonalized by the Fourier transform within that subspace. 

In one dimension the momentum eigenstates are represented by the $L$th roots of unity $e^{2\pi i n/L}$. The corresponding eigenvalue of the hopping term is readily shown to be $2J_r\cos(2\pi n/L).$ In $d$-dimensions the first-order corrections to the energy eigenvalues are
\begin{equation}
    E_{n_i} = 2dJ_r \cos(2\pi n_i/L).
\end{equation}
For large $L$ the first-order correction to the energy gap is therefore $-2J_r$;
\begin{equation}
    \Delta_E = 1 -2dJ_r + \mathcal{O}(J_r^2).
\end{equation}
We now consider the first-order correction to the weak-coupling ground state. Since $\ket{\Omega(0)}$ has zero momentum it can only couple to a state which also has zero momentum. The hopping term is nonzero only in the subspace with an odd number of triplet states, so it does not contribute. The pair creation term couples the weak-coupling vacuum to the subspace of the second-excited state manifold with a $\ket{p,m}$,$\ket{p,-m}$ pair on adjacent sites and zero momentum in each spatial direction (call this manifold $D$). 

The first-order correction is
\begin{equation}
    \ket{\Omega (0)^{(1)}} = -J_r \sum_{k \in D} \frac{\braket{k|H_p|\Omega(0)}}{2}\ket{k}.
\end{equation}
Note that $H_p\ket{\Omega(0)}$ is the sum of all states with adjacent $\ket{p,m}$,$\ket{p,-m}$ pairs. An orthonormal basis of $D$ is the equal superposition of $\ket{p,m}$,$\ket{p,-m}$ pairs for each of the $3$ choices of $m$. This gives an overlap of
\begin{equation*}
    \braket{k|H_p|\Omega(0)} = \sqrt{L^d},
\end{equation*}
so that
\begin{equation}
    \ket{\Omega (0)^{(1)}} = -\frac{J_r\sqrt{L^d}}{2} \sum_{k \in D}\ket{k} = -\frac{J_r}{2}H_p\ket{\Omega(0)}.
\end{equation}
We can now calculate the overlap between the ground state at finite $J_r$ and that at $J_r = 0$. The normalized finite-coupling ground state is (let $-\frac{J_r}{2}H_p\ket{\Omega(0)} = \ket{C}$)
\begin{equation}
    \ket{\Omega(J_r)} = \frac{\ket{\Omega(0)} + \ket{C}}{\sqrt{1+3J_r^2L^d/4}},
\end{equation}
so that
\begin{equation}
    \braket{\Omega(0)|\Omega(J_r)} = \left(1+3J_r^2L^d/4\right)^{-1/2}.
\end{equation}
Next, to calculate the energy gap to $\mathcal{O}(J_r^2)$, we need to obtain the first-order corrections to the first-excited states. The contributions here are from the third-excited states, which couple to the former via the pair annihilation operator. Again, momentum must be preserved. Now, however, there is a large degeneracy owing to where we decide to create the pair (for simplicity we imagine the state as having an excitation at one spatial site, create a pair somewhere else on the lattice, and extrapolate to the full state using the translation operator).  

\begin{figure}
    \centering
    \includegraphics[width=240px]{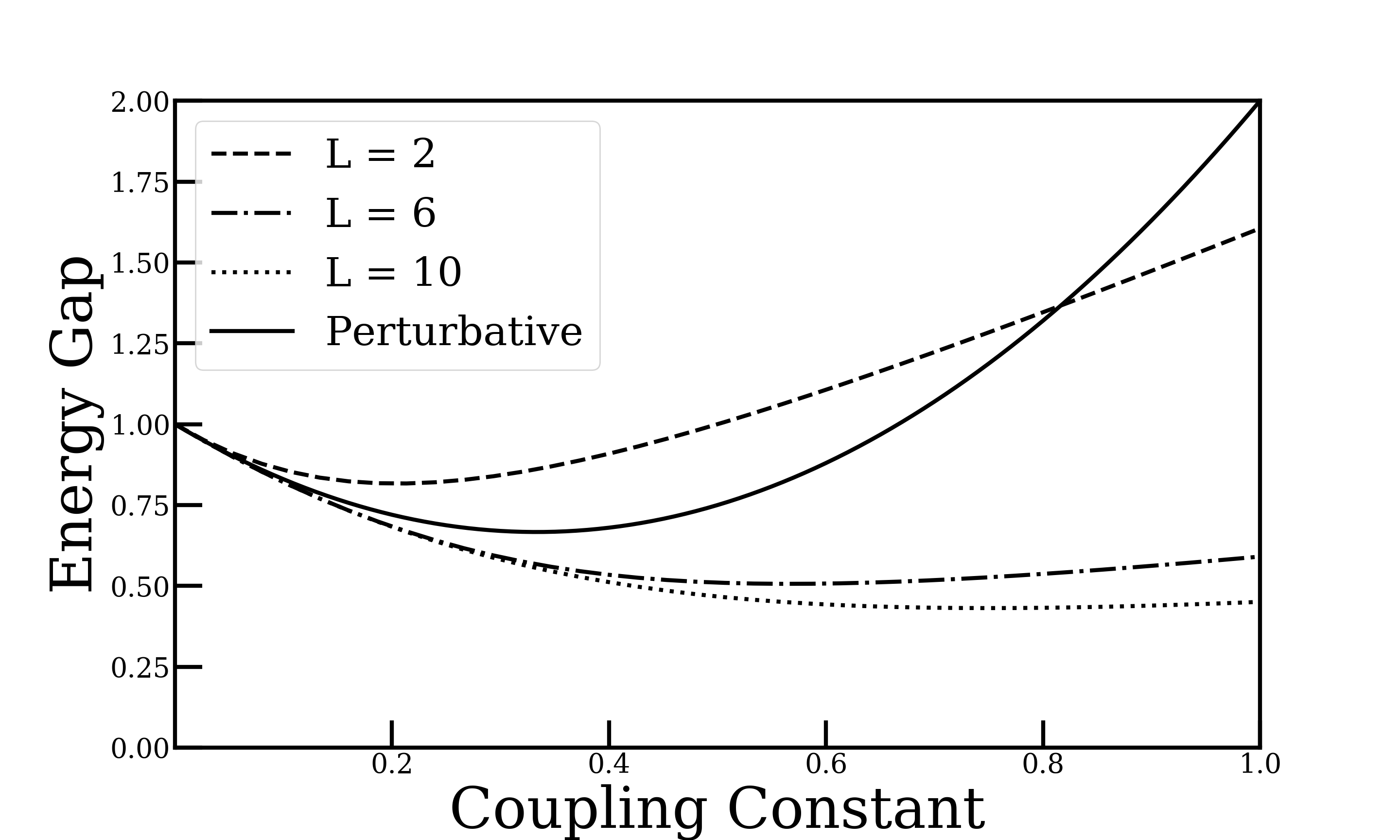}
    \caption{Ground state energy gap as a function of coupling constant in one spatial dimension for various lattice sizes, compared to perturbative result. The energy gap is nearly linear in coupling for larger values of $J_r$ (not shown). The spectrum appears to converge to something which is gapped at all values of $J_r$ as $L$ is increased.}
    \label{fig:1dgap}
\end{figure}
\begin{figure}
    \centering
    \includegraphics[width=240px]{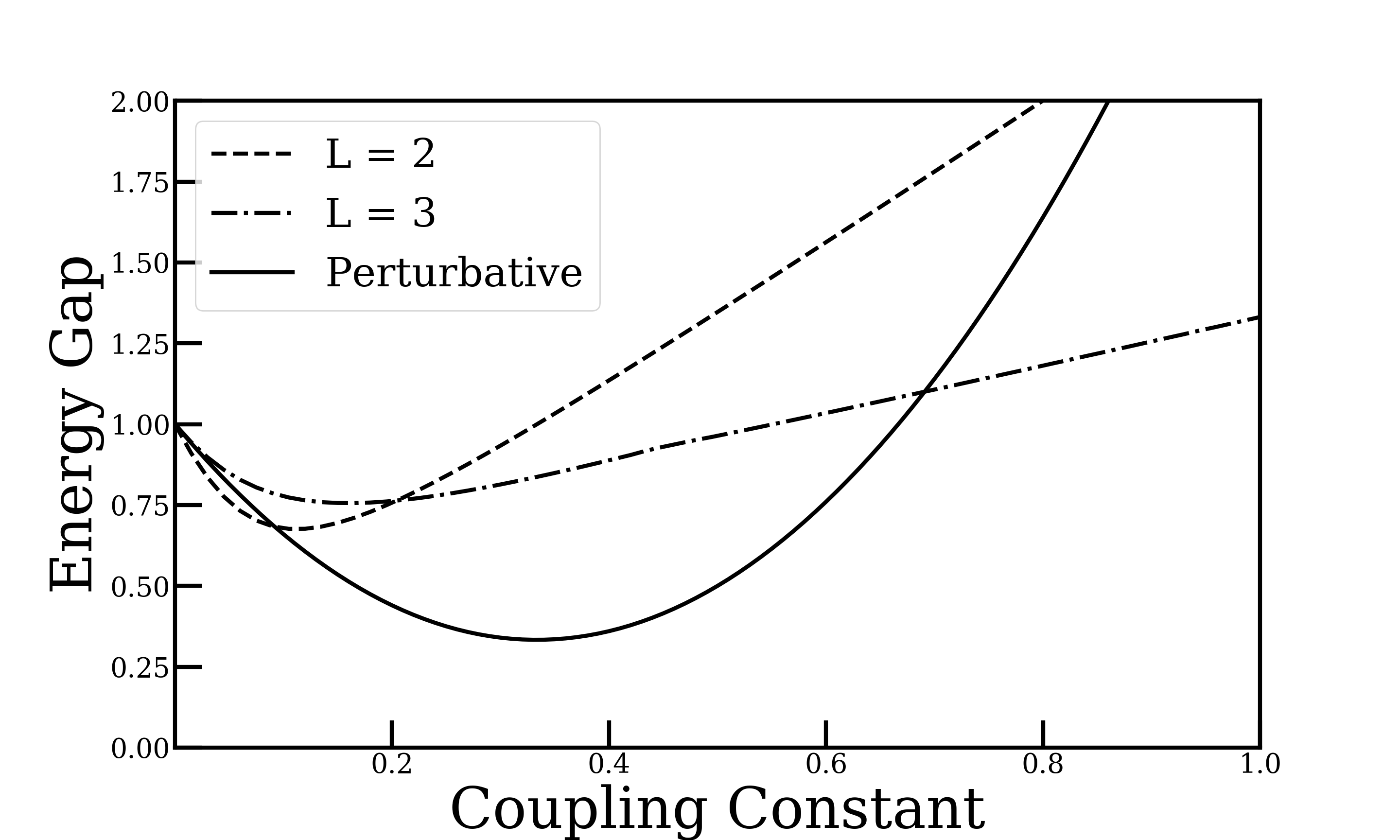}
    \caption{Ground state energy gap as a function of coupling constant in two spatial dimensions for $L = 2,3$, compared to perturbative result. The energy gap is nearly linear in coupling for larger values of $J_r$ (not shown).}
    \label{fig:2dgap}
\end{figure}

Let $\ket{k,m}_1$ be a first-excited state with momentum $k$ and $M = m$. We can create a pair at any of $L^d - 2d$ sites (call the site $r$), and we have $3$ options for the $m$-value of that pair (call this $m'$), so that $\ket{k,m}_1$ couples to exactly $(3L^d - 6d)$ third-excited states (call these $\ket{k,m,r,m'}$). This gives an overlap
\begin{equation}
    \braket{k,m|_1 H_p |k,m,r,m'} = 1, 
\end{equation}
and
\begin{equation}
    \ket{k,m^{(1)}}_1 = -J_r\sum_{r,m'} \frac{1}{2}\ket{k,m,r,m'} = -\frac{J_r}{2} H_p \ket{k,m}_1
\end{equation}
We find second-order corrections to the ground and first excited state energies of
\begin{equation}
  \begin{split}
    E_0^{(2)} &= -3J_r^2L^d/2, \\
    E_1^{(2)} &= -3J_r^2(L^d - 2d)/2,
    \end{split}
\end{equation}
so that the energy gap to second order in the coupling is
\begin{equation}
    \Delta E = 1 - 2dJ_r + 3dJ_r^2 + \mathcal{O}(J_r^3).
\end{equation}
We find reasonable agreement between this result and that obtained via exact diagonalization on finite lattices (see Figures~\ref{fig:1dgap} and ~\ref{fig:2dgap}).

\section{General Symmetric Model}
\label{sec:general}
In this section we provide a complete characterization of all translation invariant nearest-neighbor Hamiltonians with two qubits per lattice site that are invariant under $SO(3)$ . This treatment serves to clarify some of the statements made in the main text. Let $g \in \text{SO(3)}$, and let $U^{\rho}_g$ be the representation of $g$ under the irrep $\rho$. The Hilbert space on each lattice site transforms as the direct sum $0 \oplus 1$. If $H$ is a Hamiltonian acting on two sites, $H$ transforms as the direct sum $0 \oplus 1 \oplus 2$, with multiplicities $2$, $3$, and $1$, respectively. 

In order to conjugate $H$ by a group element $g$ on each site, we apply $(U^{0}_g \oplus U^{1}_g) \otimes (U^{0}_g \oplus U^{1}_g)$. This is unitarily equivalent to the following $16 \times 16$ matrix;
\begin{equation*}
    U_g = \begin{bmatrix}
    U^{2}_g & 0 & 0 & 0 & 0 & 0 \\
    0 & U^{1}_g & 0 & 0 & 0 & 0 \\
    0 & 0 & U^{1}_g & 0 & 0 & 0 \\
    0 & 0 & 0 & U^{1}_g & 0 & 0 \\
    0 & 0 & 0 & 0 & U^{0}_g & 0 \\
    0 & 0 & 0 & 0 & 0 & U^{0}_g \\
    \end{bmatrix}.
\end{equation*}
If $H$ is $SO(3)$ -invariant, then $U_g H U_g^{\dagger} = H$ for all $g \in \text{SO(3)}$. Writing $H$ in the same basis as that which block diagonalizes $U_g$, we can think of $H$ in block-form, where each block defines a linear map between representations $j$ and $j'$. According to Schur's lemma, if a block commutes with the group action it is either $0$ or invertible. Thus the blocks coupling \textit{different} representations are $0$, and the blocks coupling equivalent representations are proportional to the identity. This implies that $H$ takes the following simple form
    \begin{equation*}
    H = \begin{bmatrix}
    aI_5 & 0 & 0 & 0 & 0 & 0 \\
    0 & bI_3 & cI_3 & dI_3 & 0 & 0 \\
    0 & eI_3 & fI_3 & gI_3 & 0 & 0 \\
    0 & hI_3 & iI_3 & jI_3 & 0 & 0 \\
    0 & 0 & 0 & 0 & k & l \\
    0 & 0 & 0 & 0 & m & n \\
    \end{bmatrix},
\end{equation*}
where $I_n$ is the $n\times n$ identity matrix. 

The most general $SO(3)$  invariant four-qubit Hamiltonian can then be written as
\begin{equation}
    H = I_5 \oplus (I_3 \otimes H_{j=1}) \oplus (H_{j=0}),
\end{equation}
where $H_{j=1}$ and $H_{j=0}$ are $3\times 3$ and $2 \times 2$ Hermitian matrices, respectively. Any additional symmetries of the Hamiltonian arise from symmetries of $H_{j=1}$ and $H_{j=0}$. 

For convenience we work in the standard basis, in which the three $l = 1$ blocks are those of the ordinary symmetry channels, $1\oplus 1$, $1\oplus 0$, and $0 \oplus 1$, in that order. The $l = 0$ blocks are the $1 \oplus 1$ and the $ 0 \oplus 0$ channels, in that order. 

\subsection{Parity}
In this context, by parity we mean ordinary spatial inversion; for example, in one-dimension site $i$ is mapped to site $L-i$. Since our general symmetric Hamiltonian is translation-invariant and nearest-neighbor, under parity it is mapped to another translation-invariant nearest-neighbor Hamiltonian. Identifying the nearest-neighbor terms on a given pair of lattice sites before and after a parity transformation, in the basis considered above the parity operator is represented as 
\begin{equation}
    P = \begin{bmatrix}
    I_5 & 0 & 0 & 0 & 0 & 0 \\
    0 & -I_3 & 0 & 0 & 0 & 0 \\
    0 & 0 & 0 & I_3 & 0 & 0 \\
    0 & 0 & I_3 & 0 & 0 & 0 \\
    0 & 0 & 0 & 0 & 1 & 0 \\
    0 & 0 & 0 & 0 & 0 & 1 \\
    \end{bmatrix}.
\end{equation}
If $H$ is $SO(3)$  invariant and $PHP^{-1} = H$, this implies that
% \begin{equation}
 \begin{align}
    H &= \begin{bmatrix}
    aI_5 & 0 & 0 & 0 & 0 & 0 \\
    0 & bI_3 & cI_3 & dI_3 & 0 & 0 \\
    0 & c^*I_3 & eI_3 & fI_3 & 0 & 0 \\
    0 & d^*I_3 & f^*I_3 & gI_3 & 0 & 0 \\
    0 & 0 & 0 & 0 & h & i \\
    0 & 0 & 0 & 0 & i^* & j \\
    \end{bmatrix} \nonumber\\ &= 
    \begin{bmatrix}
    aI_5 & 0 & 0 & 0 & 0 & 0 \\
    0 & bI_3 & -dI_3 & -cI_3 & 0 & 0 \\
    0 & -d^*I_3 & gI_3 & f^*I_3 & 0 & 0 \\
    0 & -c^*I_3 & fI_3 & eI_3 & 0 & 0 \\
    0 & 0 & 0 & 0 & h & i \\
    0 & 0 & 0 & 0 & i^* & j \\
    \end{bmatrix}.
    \end{align}
%\end{equation}
Without imposing parity there are $14$ real degrees of freedom. Under parity, four of these are removed ($c = -d$, $e = g$, $f = f^*$). 

The most general $SO(3)$  and parity-invariant Hamiltonian is then
\begin{equation}\label{general}
    H = \begin{bmatrix}
    aI_5 & 0 & 0 & 0 & 0 & 0 \\
    0 & bI_3 & cI_3 & -cI_3 & 0 & 0 \\
    0 & c^*I_3 & dI_3 & eI_3 & 0 & 0 \\
    0 & -c^*I_3 & eI_3 & dI_3 & 0 & 0 \\
    0 & 0 & 0 & 0 & f & g \\
    0 & 0 & 0 & 0 & g^* & h \\
    \end{bmatrix}.
\end{equation} 
Lastly, imposing time-reversal symmetry, we derive that $c = 0, g= g^*$. Using our freedom to subtract an overall constant, The full $SO(3)$ , P, and T-symmetric Hamiltonian (with $6$ real degrees of freedom) is
\begin{equation}\label{tsym}
    H = \begin{bmatrix}
    aI_5 & 0 & 0 & 0 & 0 & 0 \\
    0 & bI_3 & 0 & 0 & 0 & 0 \\
    0 & 0 & dI_3 & eI_3 & 0 & 0 \\
    0 & 0 & eI_3 & dI_3 & 0 & 0 \\
    0 & 0 & 0 & 0 & f & g \\
    0 & 0 & 0 & 0 & g & 0 \\
    \end{bmatrix}.
\end{equation} 
We would like to connect this general model to our original qubit-regularized Hamiltonian. From now on, we neglect the cumbersome $I_n$ in each block. The terms in our original Hamiltonian take the form (setting $\mu = 0$)
\begin{align}
    H^h & = \begin{bmatrix}
    0 & 0 & 0 & 0 & 0 & 0\\
    0 & 0 & 0 & 0 & 0 & 0\\
    0 & 0 & 0 & 1 & 0 & 0\\
    0 & 0 & 1 & 0 & 0 & 0\\
    0 & 0 & 0 & 0 & 0 & 0\\
    0 & 0 & 0 & 0 & 0 & 0\\
    \end{bmatrix} \\
    H^p & = \begin{bmatrix}
    0 & 0 & 0 & 0 & 0 & 0\\
    0 & 0 & 0 & 0 & 0 & 0\\
    0 & 0 & 0 & 0 & 0 & 0\\
    0 & 0 & 0 & 0 & 0 & 0\\
    0 & 0 & 0 & 0 & 0 & 1\\
    0 & 0 & 0 & 0 & 1 & 0\\
    \end{bmatrix} \\
    H_1 &= \begin{bmatrix}
    2 & 0 & 0 & 0 & 0 & 0 \\
    0 & 1 & 0 & 0 & 0 & 0 \\
    0 & 0 & 1 & 0 & 0 & 0 \\
    0 & 0 & 0 & 1 & 0 & 0 \\
    0 & 0 & 0 & 0 & 0 & 0 \\
    0 & 0 & 0 & 0 & 0 & 0 \\
    \end{bmatrix}.
\end{align}
We get back to the most general model by adding three additional couplings
\begin{eqnarray}
    H^X & = \begin{bmatrix}
    1 & 0 & 0 & 0 & 0 & 0\\
    0 & -1 & 0 & 0 & 0 & 0\\
    0 & 0 & 0 & 0 & 0 & 0\\
    0 & 0 & 0 & 0 & 0 & 0\\
    0 & 0 & 0 & 0 & 1 & 0\\
    0 & 0 & 0 & 0 & 0 & 0\\
    \end{bmatrix} \\
    H^= &= \begin{bmatrix}
    0 & 0 & 0 & 0 & 0 & 0\\
    0 & 0 & 0 & 0 & 0 & 0\\
    0 & 0 & 0 & 0 & 0 & 0\\
    0 & 0 & 0 & 0 & 0 & 0\\
    0 & 0 & 0 & 0 & 1 & 0\\
    0 & 0 & 0 & 0 & 0 & 0\\
    \end{bmatrix} \\
    H^{sm} &= \begin{bmatrix}
    0 & 0 & 0 & 0 & 0 & 0\\
    0 & 0 & 0 & 0 & 0 & 0\\
    0 & 0 & 1 & 0 & 0 & 0\\
    0 & 0 & 0 & 1 & 0 & 0\\
    0 & 0 & 0 & 0 & 0 & 0\\
    0 & 0 & 0 & 0 & 0 & 0\\
    \end{bmatrix}
\end{eqnarray}
Including all the couplings, we obtain the Hamiltonian
\begin{equation}\label{eqn:general}
    \setlength{\arraycolsep}{0pt}
    H = \begin{bmatrix}
    2J_t+J_X & 0 & 0 & 0 & 0 & 0 \\
    0 & J_t-J_X & 0 & 0 & 0 & 0 \\
    0 & 0 & J_t+J_{sm} & J_h & 0 & 0 \\
    0 & 0 & J_h & J_t+J_{sm} & 0 & 0 \\
    0 & 0 & 0 & 0 & J_= + J_X & J_p \\
    0 & 0 & 0 & 0 & J_p & 0 \\
    \end{bmatrix}.
\end{equation}

% \onecolumngrid

\end{document}